\documentclass{aa}
\usepackage{graphicx}
\usepackage{txfonts}
\usepackage{longtable}
\usepackage{lscape}

\usepackage{longtable}

\begin{document}
   \title{Ultra-cool dwarfs: new discoveries, proper motions, and improved spectral
   typing from SDSS and 2MASS photometric colors}

   \author{Z.H. Zhang\inst{1,3,2}, R. S. Pokorny\inst{1}, H. R. A.
   Jones\inst{2}, D. J. Pinfield\inst{2}, P.S. Chen\inst{1},  Z. Han\inst{1},  \\
   D. Chen\inst{1,2}, M. C. G\'{a}lvez-Ortiz\inst{2} \and B. Burningham\inst{2}}

    \titlerunning{New ultra-cool dwarfs from the SDSS}

    \authorrunning{Z.H. Zhang et al.}

   \offprints{Z.H. Zhang}

   \institute{National Astronomical Observatories/Yunnan
    Observatory, Chinese Academy of Sciences, Kunming 650011,
    China\\
    \email{zenghuazhang@hotmail.com}
     \and
    Centre for Astrophysics Research, Science and Technology Research
    Institute, University of Hertfordshire, College Lane, Hatfield AL10 9AB,
    U.K.\\
    \email{h.r.a.jones@herts.ac.uk}
    \and
    Graduate School of Chinese Academy of Sciences, Beijing 100049, China\\}

   \date{Received June 4, 2008; accepted January 14, 2009}


  \abstract
   {}
   {We try to identify ultra-cool dwarfs from the seventh Data Release of the Sloan
   Digital Sky Survey (SDSS DR7) with SDSS \emph{i-z} and \emph{r-z} colors. We also
   obtain proper motion data from SDSS, 2MASS, and UKIDSS and improve spectral
   typing from SDSS and 2MASS photometric colors.}
   {We selected ultra-cool dwarf candidates from the SDSS DR7 with new photometric
   selection criteria, which are based on a parameterization study of known
   L and T dwarfs. The objects are then cross-identified with the Two
    Micron All Sky Survey and the Fourth Data Release of the
    UKIRT Infrared Deep Sky Survey (UKIDSS DR4). We derive proper motion constraints by
    combining SDSS, 2MASS, and UKIDSS positional information. In this way
    we are able to assess, to some extent, the credence of our sample
    using a multi epoch approach, which complements spectroscopic
    confirmation. Some of the proper motions are affected by short
    baselines, but, as a general tool, this method offers great
    potential to confirm faint L dwarfs as UKIDSS coverage increases.
    In addition we derive updated color-spectral type relations
    for L and T dwarfs with SDSS and 2MASS magnitudes.}
   {We present 59 new nearby M and L dwarfs selected from the imaging
    catalog of the SDSS DR7, including proper motions and spectral types
    calculated from the updated color-spectral type relations. and obtain
    proper motions from SDSS, 2MASS, and UKIDSS for all of our objects.}
   {}


   \maketitle
%

\section{Introduction}
Brown dwarfs occupy the mass range between the lowest mass stars and
the highest mass planets. The central temperature of a brown dwarf
is not high enough to achieve stable hydrogen burning like a star,
but all brown dwarfs will undergo short periods of primordial
deuterium burning very early in their evolution. Since the first
discovery of an L dwarf  (GD165 B; Becklin \& Zuckerman
\cite{bec88}) and a T dwarf (Gl229 B; Nakajima et al. \cite{nak95}),
the projects searching for brown dwarfs have involved a number of
large scale surveys, for example, the Deep Near-Infrared Survey
(DENIS; Epchtein et al. \cite{epc97}), the Two Micron All Sky Survey
(2MASS; Skrutskie et al. \cite{skr06}) and the Sloan Digital Sky
Survey (SDSS; York et al. \cite{yor00}; Adelman-MaCarthy et al.
\cite{ade08}). 554 L dwarfs and 145 T dwarfs have been found in
large scale sky surveys in the last decade (by January 2009, see,
DwarfsArchives.org for a full list). 31 of L or T dwarfs have been
found in DENIS (e.g. Delfosse et al. \cite{del97}), 185 in SDSS
(e.g. Fan et al. \cite{fan00}; Geballe et al. \cite{geb02}; Hawley
et al. \cite{haw02}; Schneider et al. \cite{sch02}; Knapp et al.
\cite{kna04}; Chiu et al. \cite{chi06}), and 368 in 2MASS (e.g.
Burgasser et al. \cite{bur99}, \cite{bur02}, \cite{bur04};
Kirkpatrick et al. \cite{kir99}, \cite{kir00}; Gizis et al.
\cite{giz00}; Cruz et al. \cite{cru03}, \cite{cru07}; Kendall et al.
\cite{ken03}, \cite{ken07a}; Looper et al. \cite{loo07}). More
recently, the UKIRT Infrared Deep Sky Survey (UKIDSS; Lawrence et
al. \cite{law07}) is beginning to be very effective in searching for
T dwarfs (Kendall et al. \cite{ken07b}; Lodieu et al.\cite{lod07};
Warren et al. \cite{war07}; Burningham et al. \cite{bur08}; Pinfield
et al. \cite{pin08}).

In this paper we report the discovery of 59 late M and L dwarfs from
the main photometric catalog of SDSS DR7.1. The photometric
selection processes are presented in section 2. The red optical
spectra of the 36 new late M and L dwarfs from SDSS are presented in
section 3. Polynomial fitting for color-spectral type relationships
are derived in section 4. The UKIDSS matches for 23 ultra-cool dwarf
candidates without SDSS spectra are presented in section 5, and
section 6 presents some further discussion.


\section{Photometric selection}
The Sloan Digital Sky Survey uses a dedicated 2.5 m telescope
located at Apache Point Observatory (APO) in New Mexico. It is
equipped with a large format mosaic CCD camera to image the sky in
five optical bands (\emph{u}, \emph{g}, \emph{r}, \emph{i},
\emph{z}), and two digital spectrographs to obtain the spectra of
galaxies, quasars and late type stars selected from the imaging data
(York et al. 2000). The SDSS DR7 imaging data covers about 8420
$deg^{2}$ of the main survey area (legacy sky), with information on
roughly 230 mil-
\begin{figure*}
\centering
\includegraphics[width=13.5cm]{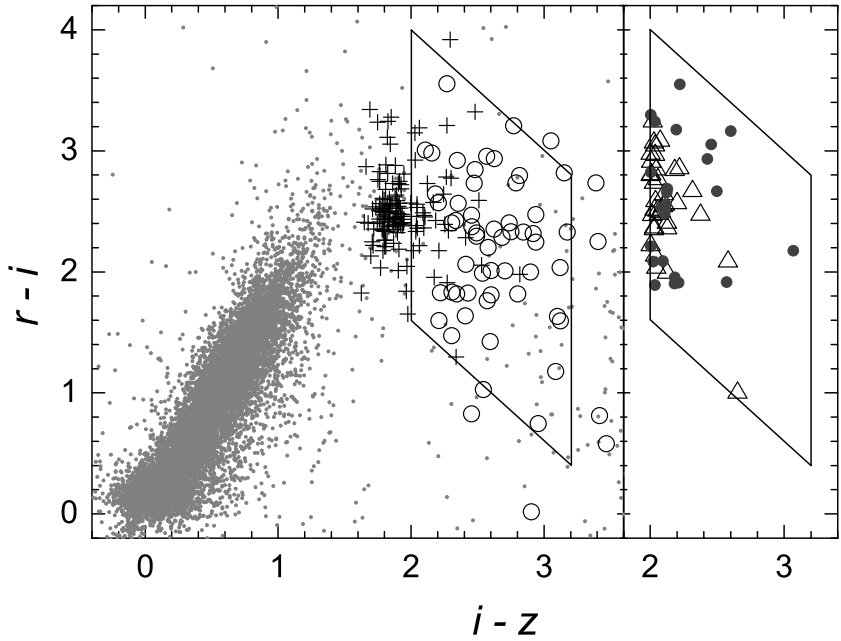}
\caption{$r-i$ vs. $i-z$ diagram for previously identified L dwarfs
(left hand panel) and new M and L dwarfs (right hand panel).
Previously identified L0-L4.5 and L5-L9.5 dwarfs are shown as
\emph{crosses} and \emph{open circles} respectively. The 36 new M
and L dwarfs with measured spectral types (from SDSS spectroscopy)
are shown as \emph{open triangles}, and the 23 new ultracool dwarf
candidates (no SDSS spectra available) that we have cross-matched in
UKIDSS DR4 are shown as \emph{filled circles}. As a comparison,
24300 point sources with $15<z<20.5$ from 10 $deg^{2}$ of SDSS
coverage are shown as \emph{dots}. A parallelogram shows the
boundary limits of our color selection.} \label{f1}
\end{figure*}
lion distinct photometric objects. The SDSS magnitude limits (95\%
detection repeatability for point sources) for the \emph{u},
\emph{g}, \emph{r}, \emph{i} and \emph{z} bands are 22.0, 22.2,
22.2, 21.3 and 20.5 respectively.

The $i-z$ color is particularly useful for L dwarf selection (as first
pioneered by Fan et al. (\cite{fan00}), and expanded on by others e.g.
via the $i$-band drop-out method; e.g. Chiu et al. \cite{chi06}). For
the cooler $T_{\rm eff}$ T dwarfs, almost all of the radiation is emitted
beyond 10000 {\AA}, and as such these objects are optically much fainter
than L dwarfs. SDSS is thus significantly less sensitive to T dwarfs than
to L dwarfs, but has the sensitivity to identify L dwarfs out to distances
well beyond 100pc.

We have made a study of L dwarf color-color parameter-space using
previously identified L and T dwarfs with photometric data available
from either SDSS or 2MASS (from DwarfsArchives.org, as of September
25, 2007). Where two spectral types are available (optical and
infrared) we used the mean average type. A total of 431 L and 84 T
dwarfs have 2MASS photometric data (\emph{J}, \emph{H}, \emph{K}),
and 193 L and 46 T dwarfs have SDSS photometric data (\emph{u},
\emph{g}, \emph{r}, \emph{i}, \emph{z}). We excluded L and T dwarfs
known to be unresolved binary systems from our study. These optical
and near-infrared parameter spaces are shown in the left-hand panels
of Figures 1 and 2, where crosses indicate early L dwarfs (L0-L4.5)
and open circles indicate mid-late L dwarfs (L5-L9.5). Using these
plots we have identified regions of color space that contain the
vast majority of mid-late L dwarfs. We chose $i-z>2$ to avoid too
much contamination from red dwarfs, although this does leads to
missing many early L dwarfs. Figure~\ref{f1} shows the selection
cuts in the $r-i$ versus $i-z$ diagram, in which a parallelogram
shows the boundary limits. Two two sloping lines show the boundary
limits $r-z=3.6$ and $r-z=6$, and the two vertical lines show
$i-z=2$ and $i-z=3.2$. Taking into account also the photometric
sensitivities of SDSS, we thus define a set of SDSS mid-late L dwarf
photometric selection criteria as follows:
\begin{eqnarray}
19 < i < 23  \\
17 < z < 20  \\
3.6  < r-z < 6 \\
2 < i-z < 3.2
\end{eqnarray}

Criterion (3) can also be written as
\begin{equation}
3.6-(i-z)<r-i<6-(i-z)
\end{equation}
in \emph{r~i~z} color-color space.

The left-hand panel of Figure 2 shows the $J-H$ vs. $H-K$ diagram
for known L dwarfs. As for Figure 1 we define a set of color
criteria to contain the majority of these L dwarfs. It is clear from
Figure 2 that the L dwarfs appear reasonably well separated from the
M dwarfs in this 2-color space (see also Burgasser et al.
\cite{bur02}), and we would thus expect to improve our sample
refinement significantly by combining our optical selection with
additional near infrared photometry. Our chosen 2MASS color
selection criteria are shown in the Figure with solid lines, and are
defined as:
\begin{eqnarray}
J-H  > & 0.5  \\
H-K  > & 0.2   \\
J-K  > & 1
\end{eqnarray}

\begin{figure*}
\centering
\includegraphics[width=18cm]{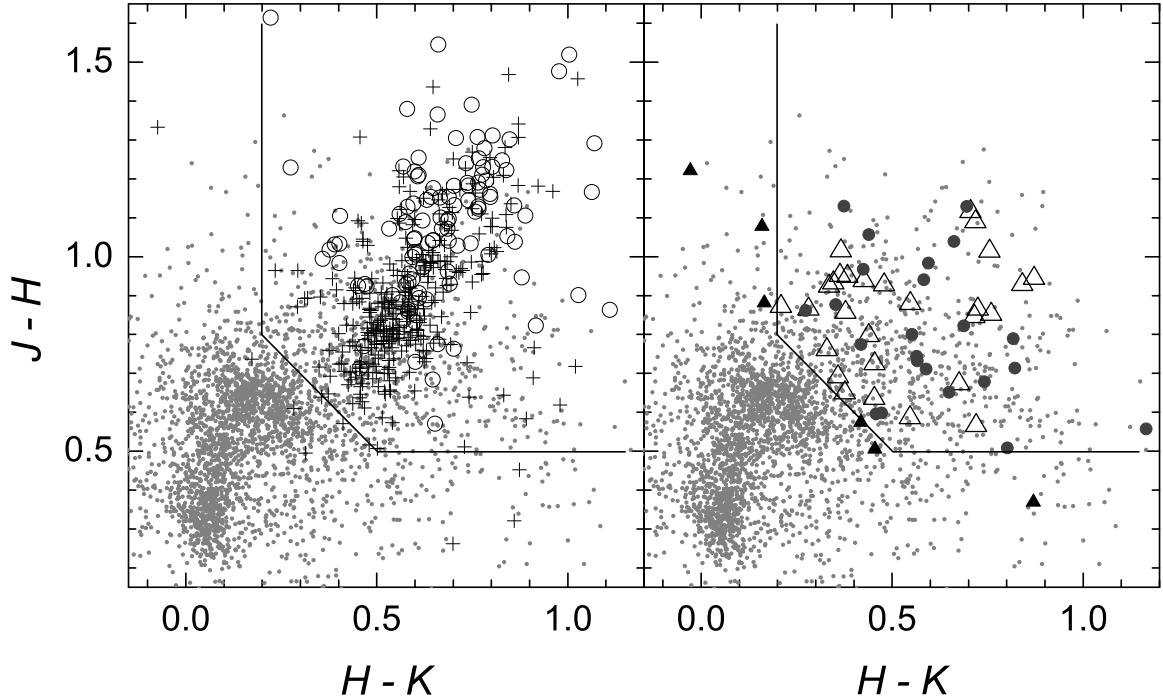}
\caption{$J-H$ vs. $H-K$ diagram for previously identified L dwarfs
(left hand panel) and new M and L dwarfs (right hand panel). Symbols
are as in Figure 1 except that six of the new spectroscopically
confirmed dwarfs (three M9, one L0 and two L1) are indicated with
filled triangles, because they lie outside our 2MASS photometric
selection criteria (solid lines). Note that the 2MASS criteria are
only applied when selecting photometric candidates, and not when
SDSS spectra are available. For comparison, the plot shows 2800
sources (dots) taken from 3.14 $deg^2$ of 2MASS sky.} \label{f2}
\end{figure*}

More than 6000 candidates survived our color and magnitude optical
selection from the main photometric catalog of SDSS DR7. Those
candidates were matched with point sources in the 2MASS catalog
(Skrutskie et al. \cite{skr06}). We used a matching radius of
6$^{\prime\prime}$ to ensure that any ultra-cool dwarfs with high
proper motion could be matched, despite possible motion over a
period of up to $\sim$8 yrs (between epochs). A total of 700 SDSS
objects were cross-matched in 2MASS. Because the imaging depth of
SDSS is deeper than that of 2MASS, many of the fainter SDSS
candidates could not be found in 2MASS. Thus some of the candidates
get wrongly matched with their nearest brighter neighbors in 2MASS.
These are removed by checking their images in both the SDSS and
2MASS databases. On closer inspection, 2MASS mis-matches are usually
very apparent, since the mis-matching objects generally have their
own SDSS counterpart that immediately rules them out as high proper
motion. As a more subtle check of the cross-matching, we assessed
the spectral energy distribution of candidates (estimated from their
SDSS and 2MASS photometry), and identified candidates that appeared
to have unusual SEDs when compared to those of known ultra-cool
dwarfs. It was assumed that these objects were also 2MASS
mis-matches, and they were removed from our selection. Tens of
wrongly matched objects were thus identified and removed from our
selection. Implementation of our near infrared color selection
criteria (for candidates that did not have SDSS spectroscopy)
allowed us to remove more than 300 objects from our sample via the
$J-H$ and $H-K$ color cuts listed in equations 6, 7 and 8.

\section{Red optical spectra from the SDSS}
The SDSS imaging data are used to select in a uniform way different
classes of objects whose spectra will be taken with the SDSS 2.5 m
telescope (York et al. \cite{yor00}). The target selection
algorithms for spectroscopic follow up are described by Stoughton et
al. (\cite{sto02}). The DR7.1 main spectroscopic data base includes
data for around 1.2 million objects, and covers 7470 $deg^{2}$. The
wavelength coverage is from 3800 to 9200 {\AA} with resolution
$\lambda/(\Delta\lambda)=1800$. The signal-to-noise ratio is better
than 4 pixel$^{-1}$ at $g=20.2$ (Adelman-McCarthy et al.
\cite{ade07}). The spectra distributed by the SDSS have been sky
subtracted and corrected for telluric absorption. The spectroscopic
data are automatically reduced by the SDSS pipeline software.

Our final selection consisted of 275 objects of which 87 were cross-
referenced with, and confirmed to be mostly L dwarfs. So 188
of these objects are new discoveries. Of these, 36 are confirmed
through SDSS optical spectra. At first we only found 28 objects with
SDSS spectra when we searched for the photometrically selected
candidates in the SDSS spectroscopic catalog. Then we searched
the spectra of SDSS color selected candidates with SDSS spectra which were removed
with 2MASS criteria (6, 7, 8). We found
six objects with SDSS spectra, three M9, one L0 and two L1
dwarfs. In Figure~\ref{f2}, we can see that the 2MASS criteria (6,
7, 8) are set for L dwarfs and some late M
early L dwarfs will be missed, including these six objects. Another two
objects were found with SDSS spectra which are faint and therefore
missed by the 2MASS survey. Table 1 lists the SDSS names, SDSS
\emph{r}, \emph{i}, \emph{z}, 2MASS \emph{J}, \emph{H}, \emph{K} and
SDSS spectral types for these 36 spectroscopically confirmed
ultracool dwarfs. Note that we also performed a cross-match with
UKIDSS DR4 (see Section 5) and that 23 objects without SDSS
spectroscopy that were measured in UKIDSS are listed in Table 2. All
remaining candidates (i.e. without SDSS spectroscopy or UKIDSS DR4
coverage) are given in Table 7 (online data).

In addition to the photometric and spectral type analysis we also
derived proper motion constraints for our new sample, using the dual
epoch coordinates from the SDSS and 2MASS data-bases and dividing
any movement between the epochs by the observational epoch
difference. Standard errors on these proper motions are calculated
using the major axes of the position error ellipses from SDSS and
2MASS, and are dominated by the 2MASS positional uncertainties.
Systematic astrometry errors between 2MASS-SDSS which have a shorter baseline,
could lead to significant
errors in calculating proper motions with coordinates and epoch
differences, and can not be ignored. To correct the systematic
offset of 2MASS-SDSS, we measured average proper motion of
reference objects around every targets for which proper motion has been
measured. Then we subtracted this average proper motion from the
measured proper motion of each corresponding target. The number of
reference objects ranged from around 100 to a few hundreds for
different targets, with reference objects being selected to have low
coordinate errors (mostly $<$0.1$^{\prime\prime}$,
$<$0.2$^{\prime\prime}$ for the candidates with fewer reference
objects). Reference objects are within 12$^{\prime}$, 15$^{\prime}$
or 20$^{\prime}$ of our targets depends on availability.
These proper motions are also given in Tables 1 and 3 and 7 (Online
Data). Their quality and accuracy is assessed in Section 5 through
comparison with additional epoch image data and measurement of
motion relative to nearby sources surrounding each ultracool dwarf
(see also columns 10 and 11 in Table 1).

\begin{figure}
   \centering
   \includegraphics[width=9cm]{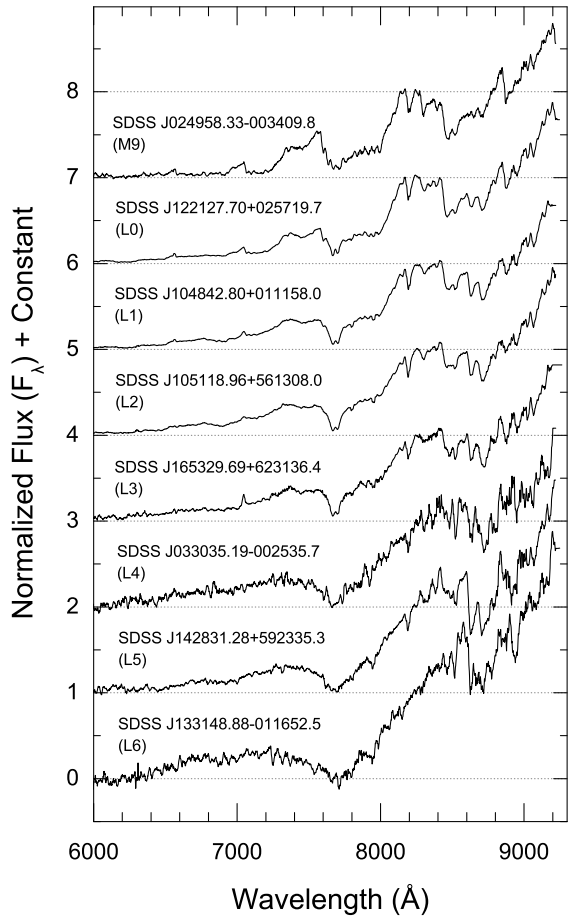}
      \caption{SDSS spectra of 8 previously found M and L dwarfs, used for spectral typing. All spectra
    have been normalized to one at 8250 {\AA} and vertically offset for clarity.}
         \label{f3}
   \end{figure}

For the 11 M9 and 25 L dwarfs with SDSS spectra. We assigned their
spectral types by comparison with the SDSS spectral sequence of
previously found M, L and T dwarfs as shown in Figure~\ref{f3}, the
dwarf classification scheme of Kirkpatrick et al. (\cite{kir99}),
SDSS spectra of M and L dwarfs published by Hawley et al.
(\cite{haw02}) and the low-mass dwarf template spectra from SDSS
(Bochanski et al. \cite{boc07}). One of the major points we have
considered in the comparison is the shape of the normalized
spectrum, such as the width of K\texttt{I} region around 7700 {\AA}
and the relative flux in the region from 8400 {\AA} to 9000 {\AA}.
Another major point is the slope of the spectrum in the band from
8700 {\AA} to 9200 {\AA}. The last criterion is the depth of
absorption lines which can be recognized in some spectra, such as
Na\texttt{I} $\lambda\lambda8183, 8195$, CrH $\lambda8661$, FeH
$\lambda8692$ and Cs\texttt{I} $\lambda\lambda8521, 8944$.
Na\texttt{I} $\lambda\lambda8183, 8195$ is a major feature of late M
dwarfs. CrH $\lambda8661$ is equal in strength to FeH $\lambda8692$
for L4 dwarfs, and stronger for L5 dwarfs. Cs\texttt{I} keeps
strengthening from L1 to L8 type (Kirkpatrick et al. \cite{kir99}).
Finally, to double check the spectral type of each spectrum, we
subtract the spectrum of a ultra-cool dwarf which has the same
spectral type and has a good quality, and find a good agreement
within our errors.

We note that our selected objects can not be giants because of the
presence of the high gravity features such as K\texttt{I},
Na\texttt{I} and FeH, which are characteristic of dwarfs (e.g.,
Bessell \cite{bes91}). Figure~\ref{f3} shows 8 spectra of previously
found M, L and T dwarfs as a comparison. SDSS J1428$+$5923 was
recently confirmed and spectral-typed using its infrared spectrum as
an L5 dwarf using the Two Micron Proper Motion (2MUPM) survey, under
the name 2MASS J14283132$+$5923354 (Schmidt et al. \cite{sch07}).
SDSS J0249$-$0034, SDSS J1048$+$0111, SDSS J1653$+$6231 and SDSS
J1331$-$0116 were classified by Hawley et al. (\cite{haw02}). SDSS
J1221$+$0257 and SDSS J1051$+$5613 were discovered by Reid et al.
(\cite{rei08}), and SDSS J0330$-$0025 was discovered by Fan et al.
(\cite{fan00}).

\begin{figure*}
   \centering
   \includegraphics[width=\textwidth]{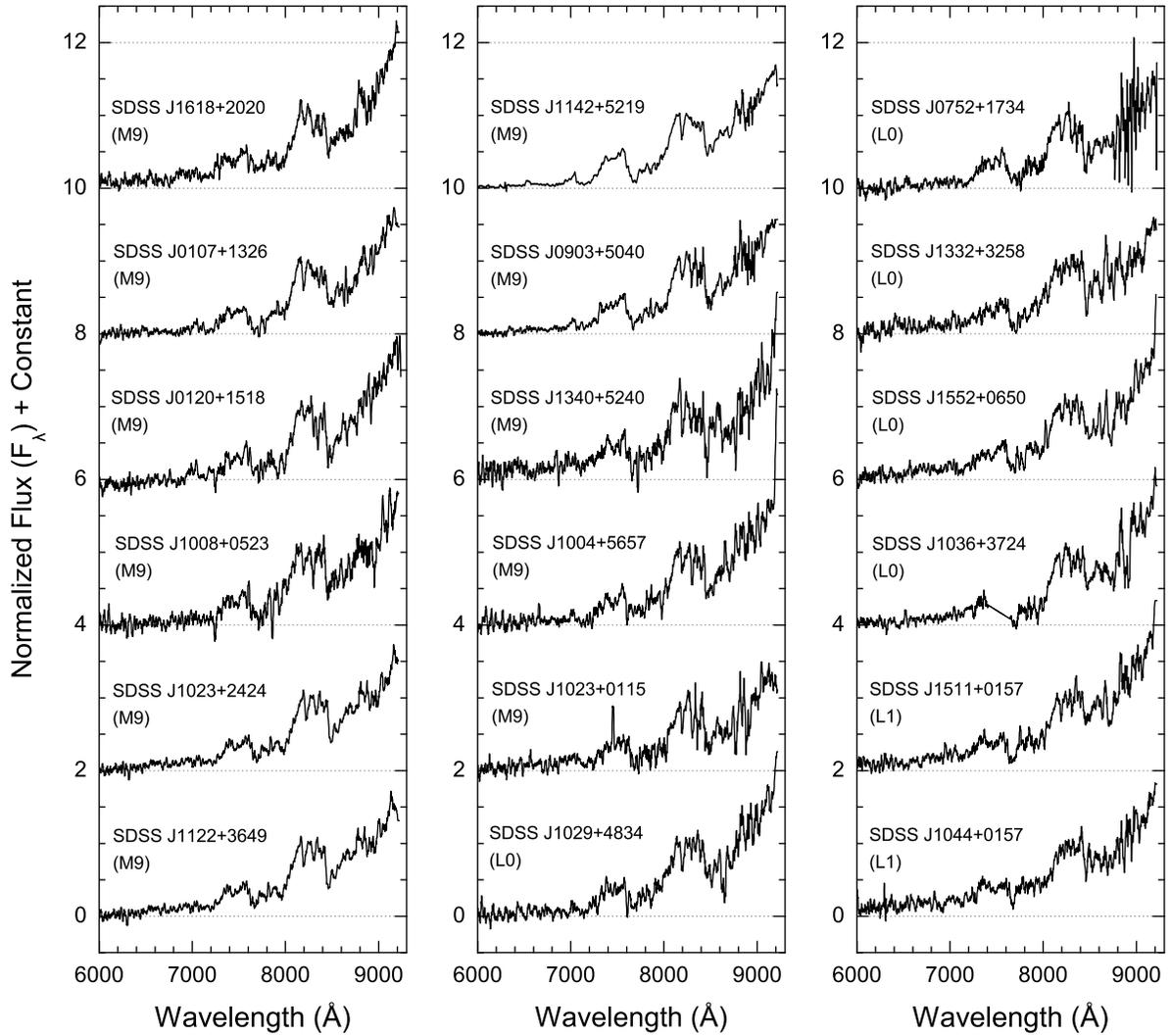}
      \caption{The SDSS spectra of new M
    and L dwarfs. Spectral types are given in brackets. The spectra are
    normalized to one at 8250 {\AA} and offset from one another for
    clarity. All the spectra are taken from the SDSS archive.}
         \label{f4}
   \end{figure*}

\begin{figure*}
   \centering
   \includegraphics[width=\textwidth]{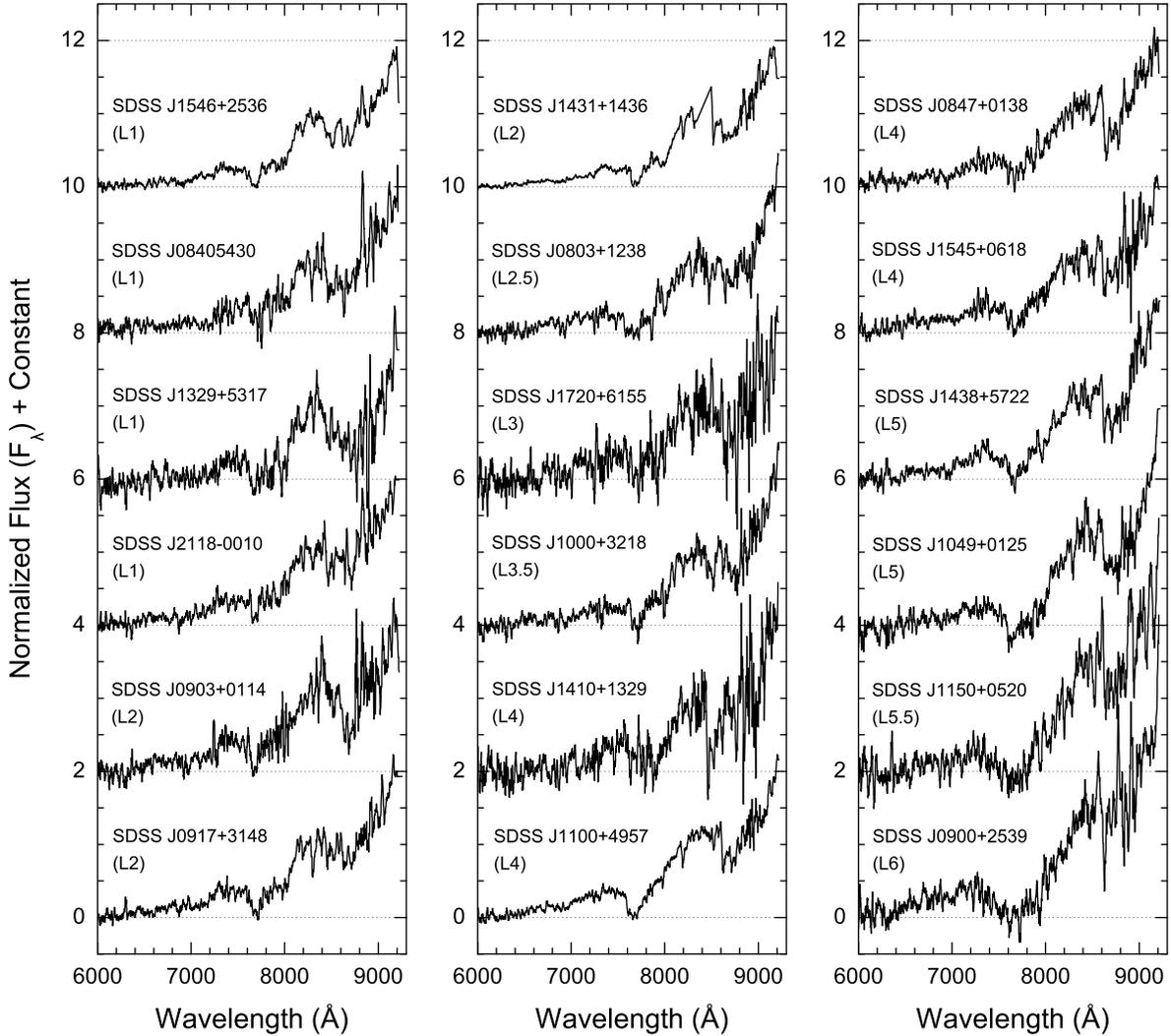}
      \caption{The SDSS spectra of new M
    and L dwarfs. Same as Figure~\ref{f4}.}
         \label{f5}
   \end{figure*}

Figures~\ref{f4} and \ref{f5} shows the SDSS spectra of the 36 M and
L dwarfs found in this work. The spectra shortward of 6000 {\AA} are
flat and noisy, and are not shown. The spectral types of our
candidates extend from M9 to L6, and the spectral typing errors are
estimated as about $\pm1\sim2$ sub-type. These SDSS spectra have
been smoothed by 11 pixels and have been normalized to one at 8250
{\AA}. There is a straight line in the spectrum of SDSS J1036$+$3724
which is an artifact. There is another artificial straight line in
the spectrum of SDSS J1431$+$1436 across 8400 {\AA}. The spectra of
SDSS J0903$+$0114, SDSS J1329$+$5317, SDSS J1410$+$1329 and SDSS
J1720$+$6155 are noisy which made it more difficult to assign their
spectral types.

\begin{landscape}
\begin{table}
 \caption{SDSS and 2MASS photometry of objects
with SDSS spectra.} \centering \label{tab1} \begin{tabular}{c c c c
c c c c l c l l l} \hline\hline
SDSS Name & SDSS \emph{r} & SDSS \emph{i} & SDSS \emph{z} & 2MASS \emph{J} & 2MASS \emph{H} & 2MASS \emph{K} & P. M.$^{\mathrm{a}}$ & P. M.$^{\mathrm{b}}$ & P. M.$^{\mathrm{c}}$ & P. M.$^{\mathrm{c}}$ & Spe. & S.T.by$^{\mathrm{d}}$ \\
 &  &  &  &  & & & ($^{\prime\prime}$yr$^{-1}$) & Angle & ($^{\prime\prime}$yr$^{-1}$) & Angle & Type & Colors    \\
\hline
~~SDSS J010718.70$+$132656.1$^{\mathrm{e}}$ & 23.24$\pm$0.31 & 20.74$\pm$0.06 & 18.68$\pm$0.04 & 16.58$\pm$0.15 & 16.21$\pm$0.19 & 15.34 ~~...~~~~ &0.51$\pm$0.13 & 289$\pm$15& 0.43 ~~...~~~~ & 107 ~~...~~~~ & M9 & ... \\
SDSS J012052.58$+$151827.3 & 23.54$\pm$0.58 & 20.69$\pm$0.08 & 18.65$\pm$0.04 & 16.39$\pm$0.13 & 15.44$\pm$0.10 & 14.57$\pm$0.12 &0.48$\pm$0.09 & 116$\pm$11& 0.13$\pm$0.09 & 209$\pm$32  & M9 & ... \\
SDSS J075256.70$+$173433.8 & 23.91$\pm$0.53 & 20.94$\pm$0.06 & 18.93$\pm$0.04 & 16.43$\pm$0.12 & 15.80$\pm$0.17 & 15.34 ~~...~~~~ &0.13$\pm$0.04 & 226$\pm$16& 0.06$\pm$0.06  & 139$\pm$136  & L0 & L1 \\
SDSS J080322.77$+$123845.3 & 23.09$\pm$0.22 & 20.73$\pm$0.05 & 18.60$\pm$0.03 & 16.31$\pm$0.08 & 15.44$\pm$0.08 & 15.23$\pm$0.11 &0.13$\pm$0.03 & 223$\pm$12& 0.28$\pm$0.23  & 199$\pm$~~3  & L2.5 & L1 \\
~~SDSS J084016.42$+$543002.1$^{\mathrm{e}}$ & 23.05$\pm$0.31 & 20.83$\pm$0.06 & 18.83$\pm$0.05 & 16.39$\pm$0.12 & 15.51$\pm$0.13 & 15.35$\pm$0.15 & ~0.51$\pm$0.81$^{\mathrm{f}}$ & 280 ... & ~1.66$\pm$1.56$^{\mathrm{f}}$  & 298$\pm$61  & L1 & L1.5 \\
SDSS J084751.48+013811.0 & 23.96$\pm$0.71 & 20.89$\pm$0.08 & 18.86$\pm$0.06 & 16.23$\pm$0.13 & 15.12$\pm$0.10 & 14.41$\pm$0.09 & ~0.66$\pm$0.47$^{\mathrm{f}}$ & 263 ... & ~1.05$\pm$1.39$^{\mathrm{f}}$  & 262$\pm$39  & L4 & L3.5 \\
SDSS J090023.68+253934.3 & 23.55$\pm$0.40 & 21.47$\pm$0.09 & 18.89$\pm$0.04 & 16.43$\pm$0.12 & 15.41$\pm$0.12 & 14.66$\pm$0.08 & 0.04$\pm$0.03 & 154$\pm$49& 0.05$\pm$0.05  & 165$\pm$12  & L6 & L4.5 \\
SDSS J090320.92+504050.6 & 23.46$\pm$0.42 & 20.55$\pm$0.05 & 18.54$\pm$0.03 & 16.42$\pm$0.12 & 15.49$\pm$0.13 & 15.15$\pm$0.14 & 0.17$\pm$0.17 & 301$\pm$73& 0.24$\pm$0.19  & 192$\pm$113  & M9 & ... \\
SDSS J090347.55+011446.0 & 23.47$\pm$0.45 & 21.07$\pm$0.08 & 18.94$\pm$0.05 & 16.45$\pm$0.14 & 15.60$\pm$0.11 & 14.89$\pm$0.13 & ~0.67$\pm$0.71$^{\mathrm{f}}$ & 205 ... & ~1.54$\pm$1.43$^{\mathrm{f}}$  & 252$\pm$71  & L2 & L2.5 \\
SDSS J091714.76+314824.8 & 23.20$\pm$0.24 & 20.85$\pm$0.04 & 18.82$\pm$0.03 & 16.30$\pm$0.10 & 15.73$\pm$0.14 & 15.01$\pm$0.13 & 0.02$\pm$0.04 & 145 ... & 0.03$\pm$0.02  & 196$\pm$73  & L2 & L.5 \\
SDSS J100016.92+321829.4 & 23.71$\pm$0.54 & 21.04$\pm$0.11 & 18.73$\pm$0.05 & 16.62$\pm$0.12 & 15.70$\pm$0.11 & 15.36$\pm$0.17 & 0.44$\pm$0.05 & 243$\pm$~~6 & 0.45$\pm$0.40  & 245$\pm$~~3  & L3.5 & L1.5 \\
SDSS J100435.88+565757.4 & 23.63$\pm$0.49 & 20.79$\pm$0.07 & 18.61$\pm$0.04 & 16.67$\pm$0.11 & 15.81$\pm$0.10 & 15.43$\pm$0.17 & 0.19$\pm$0.12 & 286$\pm$38 & ...$^{\mathrm{g}}$  & ~~~~~~...  & M9 & ... \\
~~SDSS J100817.07+052312.9$^{\mathrm{e}}$ & 23.79$\pm$0.46 & 21.02$\pm$0.06 & 18.97$\pm$0.04 & 16.76$\pm$0.16 & 16.19$\pm$0.19 & 15.77$\pm$0.28 & 0.07$\pm$0.18 & 106 ... & 0.30$\pm$0.30  & 142$\pm$~~5  & M9 & ... \\
SDSS J102316.59+011549.6 & 23.20$\pm$0.41 & 21.07$\pm$0.09 & 19.03$\pm$0.05 & 16.76$\pm$0.16 & 15.91$\pm$0.18 & 15.15$\pm$0.17 & 0.28$\pm$0.28 & 272 ... & 1.50$\pm$1.23  & 227$\pm$98  & M9 & ... \\
SDSS J102356.38+242430.3 & 22.59$\pm$0.17 & 21.59$\pm$0.12 & 18.94$\pm$0.05 &  ... &  ... &  ... & ... & ~~~~~~... & ... & ~~~~~~... & M9 & ... \\
SDSS J102947.68+483412.2 & 23.68$\pm$0.49 & 20.89$\pm$0.07 & 18.88$\pm$0.05 & 16.76$\pm$0.19 & 16.04$\pm$0.21 & 15.59$\pm$0.24 & 0.45$\pm$0.18 & 237$\pm$24& 0.23$\pm$0.16  & 112$\pm$82  & L0 & L2 \\
SDSS J103602.44+372448.5 & 22.75$\pm$0.19 & 20.76$\pm$0.06 & 18.63$\pm$0.04 & 16.42$\pm$0.12 & 15.54$\pm$0.13 & 14.99$\pm$0.11 & 0.10$\pm$0.03 & 176$\pm$17& 0.12$\pm$0.11  & 153$\pm$48  & L0 & L1 \\
SDSS J104407.47+015742.0 & 23.23$\pm$0.32 & 20.87$\pm$0.07 & 18.84$\pm$0.05 & 16.59$\pm$0.16 & 15.65$\pm$0.13 & 15.22$\pm$0.20 & 2.51$\pm$1.22 & 113$\pm$29& 1.46$\pm$1.08  & 168$\pm$65  & L1 & L0.5 \\
SDSS J104922.45+012559.2 & 23.85$\pm$0.59 & 21.00$\pm$0.09 & 18.78$\pm$0.04 & 15.88$\pm$0.07 & 14.95$\pm$0.07 & 14.11$\pm$0.06 & 0.07$\pm$0.14 & ~~21 ... & 0.13$\pm$0.16  & 200$\pm$23  & L5 & L5 \\
SDSS J110009.62+495746.5 & 22.57$\pm$0.19 & 20.08$\pm$0.03 & 18.00$\pm$0.02 & 15.28$\pm$0.04 & 14.19$\pm$0.04 & 13.47$\pm$0.03 & 0.25$\pm$0.03 & 241$\pm$~~8& 0.23$\pm$0.20  & 230$\pm$13  & L4 & L4 \\
SDSS J112242.26+364928.6 & 23.29$\pm$0.31 & 20.75$\pm$0.05 & 18.65$\pm$0.04 & 16.54$\pm$0.11 & 15.61$\pm$0.11 & 15.13$\pm$0.11 &0.07$\pm$0.03 & 106$\pm$27& 0.04$\pm$0.03  & 154$\pm$84  & M9 & ... \\
~~SDSS J114201.95+521917.0$^{\mathrm{e}}$ & 21.71$\pm$0.08 & 19.13$\pm$0.02 & 17.09$\pm$0.01 & 15.08$\pm$0.05 & 14.57$\pm$0.07 & 14.12$\pm$0.05 &0.20$\pm$0.05 & 219$\pm$13& ...$^{\mathrm{g}}$ & ~~~~~~... & M9 & ... \\
SDSS J115013.17+052012.3 & 23.75$\pm$0.39 & 21.28$\pm$0.07 & 18.91$\pm$0.04 & 16.25$\pm$0.14 & 15.46$\pm$0.14 & 15.02$\pm$0.17 &0.72$\pm$0.24 & 260$\pm$20& ...$^{\mathrm{g}}$ & ~~~~~~... & L5.5 & L3  \\
~~SDSS J132926.03+531733.9$^{\mathrm{e}}$ & 23.84$\pm$0.50 & 20.87$\pm$0.06 & 18.84$\pm$0.04 & 16.82$\pm$0.20 & 15.59$\pm$0.17 & 15.62$\pm$0.24 &0.19$\pm$0.12 & 177$\pm$40& 0.05$\pm$0.05  & 152$\pm$83  & L1 & L1 \\
~~SDSS J133257.73+325813.1$^{\mathrm{e}}$ & 23.20$\pm$0.25 & 20.73$\pm$0.04 & 18.72$\pm$0.03 & 16.57$\pm$0.11 & 15.49$\pm$0.11 & 15.33$\pm$0.12 &0.15$\pm$0.04 & 224$\pm$15& 0.12$\pm$0.12  & 218$\pm$32  & L0 & L0.5 \\
SDSS J134025.14+524505.0 & 23.71$\pm$0.55 & 21.14$\pm$0.08 & 18.94$\pm$0.04 & 16.69$\pm$0.15 & 15.67$\pm$0.14 & 15.31$\pm$0.12 &0.21$\pm$0.11 & ~~75$\pm$33& 0.15$\pm$0.16  & 159$\pm$81  & M9 & ... \\
SDSS J141011.14+132900.8 & 24.11$\pm$0.49 & 21.03$\pm$0.07 & 18.96$\pm$0.05 & 16.85$\pm$0.14 & 15.89$\pm$0.15 & 15.53$\pm$0.19 &0.18$\pm$0.06 & 279$\pm$21& 0.17$\pm$0.11  & 276$\pm$~~5  & L4 & ... \\
SDSS J143130.77+143653.4 & 22.25$\pm$0.12 & 19.69$\pm$0.02 & 17.58$\pm$0.02 & 15.15$\pm$0.04 & 14.50$\pm$0.05 & 14.13$\pm$0.06 &0.45$\pm$0.02 & 259$\pm$~~3& 0.45$\pm$0.37  & 261$\pm$~~3  & L2 & L1.5 \\
SDSS J143832.63+572216.9 & 23.45$\pm$0.40 & 20.81$\pm$0.06 & 18.71$\pm$0.04 & 15.96$\pm$0.07 & 15.10$\pm$0.08 & 14.37$\pm$0.06 &0.20$\pm$0.05 & ~~86$\pm$13& 0.12$\pm$0.14  & ~~75$\pm$43  & L5 & L3.5 \\
SDSS J151136.24+353511.4 & 23.45$\pm$0.35 & 20.72$\pm$0.05 & 18.69$\pm$0.03 & 16.29$\pm$0.10 & 15.62$\pm$0.11 & 14.95$\pm$0.12 &0.06$\pm$0.04 & 179$\pm$48& 0.16$\pm$0.13  & 133$\pm$70  & L1 & L1 \\
SDSS J154502.87+061807.8 & 23.91$\pm$0.42 & 20.87$\pm$0.05 & 18.83$\pm$0.04 & 16.37$\pm$0.11 & 15.60$\pm$0.12 & 15.28$\pm$0.16 &0.21$\pm$0.08 & 282$\pm$23& 0.20$\pm$0.14  & 263$\pm$31  & L4 & L1 \\
SDSS J154628.38+253634.3 & 22.66$\pm$0.14 & 20.15$\pm$0.03 & 18.12$\pm$0.02 & 15.76$\pm$0.07 & 15.07$\pm$0.08 & 14.71$\pm$0.08 &0.51$\pm$0.03 & 286$\pm$~~4& 0.34$\pm$0.30  & 283$\pm$13  & L1 & L0.5 \\
SDSS J155215.38+065041.5 & 24.23$\pm$0.53 & 20.99$\pm$0.06 & 18.97$\pm$0.04 & 16.78$\pm$0.15 & 15.91$\pm$0.14 & 15.63$\pm$0.21 &0.01$\pm$0.09 & 264 ... & 0.18$\pm$0.18  & 294$\pm$11  & L0 & L0.5\\
SDSS J161840.27+202045.6 & 23.40$\pm$0.26 & 20.60$\pm$0.04 & 18.59$\pm$0.03 & 16.15$\pm$0.09 & 15.20$\pm$0.07 & 14.81$\pm$0.09 &0.15$\pm$0.04 & ~~35$\pm$14& 0.04$\pm$0.04  & 205$\pm$97  & M9 & ... \\
SDSS J172006.69+615537.7 & 23.35$\pm$0.52 & 21.32$\pm$0.13 & 19.27$\pm$0.08 &  ... &  ... &  ... & ... & ~~~~~~... & ... & ~~~~~~... & L3 & ... \\
SDSS J211846.77$-$001044.6 & 23.28$\pm$0.30 & 20.80$\pm$0.06 & 18.73$\pm$0.04 & 16.20$\pm$0.11 & 15.62$\pm$0.16 & 15.07$\pm$0.13 & 0.16$\pm$0.19 & ~~18 ... & 0.29$\pm$0.27  & 223$\pm$103  & L1 & L1\\
\hline
\end{tabular}
\begin{list}{}{}
\item
Note: SDSS magnitude limits (95\% detection repeatability for point
sources) for \emph{r}, \emph{i} and \emph{z} bands are 22.2, 21.3
and 20.5 respectively.\\
\item[$^{\mathrm{a}}$]2MASS-SDSS data-base proper motions - found by dividing the difference between
the 2MASS and SDSS coordinates (from the respective databases) by
the observational epoch difference.
Standard errors are calculated using the major axes of the position error ellipses from 2MASS and SDSS.\\
\item[$^{\mathrm{b}}$]Error ellipses of 2MASS and SDSS overlap for some
objects, for which position angle errors are not meaningful.\\
\item[$^{\mathrm{c}}$]2MASS-SDSS relative proper motions - found by specifically measuring the relative
movement of the ultracool dwarfs with respect to nearby reference
objects in the 2MASS and SDSS images. \item[$^{\mathrm{d}}$]Spectral
types are based on the relationship between spectral type and SDSS
and 2MASS colors (\emph{i-z}, \emph{i-J}, \emph{i-H}, \emph{i-K} are
used largely and
\emph{r-i}, \emph{z-J}, \emph{z-H}, \emph{z-K} with less weight). \\
\item[$^{\mathrm{e}}$]Objects with SDSS spectra which do not accord with
the 2MASS color criteria (6, 7, 8).
\item[$^{\mathrm{f}}$]Objects only have a baseline of $\sim$3 months.
\item[$^{\mathrm{g}}$]We do not measure their proper motions for they are very faint in 2MASS images or only have a few nearby reference objects. \\
\end{list}
\end{table}
\end{landscape}



\begin{landscape}
\begin{table}
 \caption{Photometric data of UKIDSS matched L dwarf
candidates} \centering \label{tab2} \begin{tabular}{c c c c c c c c
c c c l} \hline\hline
SDSS Name & SDSS \emph{r} & SDSS \emph{i} & SDSS \emph{z} & 2MASS \emph{J} & 2MASS \emph{H} & 2MASS \emph{K} & UKIDSS \emph{Y} & UKIDSS \emph{J} & UKIDSS \emph{H} & UKIDSS \emph{K} & Sp.Type$^{\mathrm{a}}$ \\
\hline
~~J004759.59$+$135332.0$^{\mathrm{b}}$ & 24.78$\pm$0.88 & 22.11$\pm$0.28 & 19.61$\pm$0.10 & 16.81$\pm$0.16 & 16.10$\pm$0.17 & 15.51$\pm$0.19 & 18.03$\pm$0.03 & 17.22$\pm$0.03 & 16.52$\pm$0.03 & 16.00$\pm$0.03 & L3.5 \\
~~J015141.04$-$005156.5$^{\mathrm{c}}$ & 22.23$\pm$0.15 & 19.70$\pm$0.03 & 17.61$\pm$0.02 & 15.10$\pm$0.05 & 14.27$\pm$0.04 & 13.59$\pm$0.05 & 16.24$\pm$0.01 & 15.02$\pm$0.00 & 14.29$\pm$0.00 & 13.65$\pm$0.00 & L2 \\
J022927.95$-$005328.5 & 24.19$\pm$0.64 & 21.54$\pm$0.11 & 19.41$\pm$0.07 & 16.49$\pm$0.10 & 15.75$\pm$0.10 & 15.18$\pm$0.14 & ... & 16.62$\pm$0.02 & 15.82$\pm$0.02 & 15.15$\pm$0.02 & L3.5 \\
J073241.77$+$264558.9 & 24.31$\pm$0.46 & 22.40$\pm$0.18 & 19.83$\pm$0.08 & 17.37$\pm$0.21 & 16.81$\pm$0.25 & 15.64$\pm$0.19 & ... & 17.55$\pm$0.02 & ... & ... & L4.5 \\
J074436.02$+$251330.5 & 24.28$\pm$0.49 & 21.23$\pm$0.07 & 18.77$\pm$0.03 & 17.17$\pm$0.25 & 16.04$\pm$0.21 & 15.66$\pm$0.22 & ... & 16.68$\pm$0.01 & ... & ... & L1 \\
J075754.16$+$221604.9 & 23.34$\pm$0.36 & 21.45$\pm$0.10 & 19.41$\pm$0.06 & 16.61$\pm$0.13 & 16.10$\pm$0.22 & 15.30$\pm$0.14 & ... & 16.87$\pm$0.01 & ... & ... & L2.5 \\
J081303.96$+$243355.9 & 23.41$\pm$0.38 & 21.32$\pm$0.08 & 19.30$\pm$0.05 & 16.67$\pm$0.12 & 15.63$\pm$0.12 & 14.97$\pm$0.09 & ... & 16.57$\pm$0.01 & ... & ... & L3 \\
J081409.45$+$260250.4 & 25.06$\pm$0.67 & 21.88$\pm$0.15 & 19.69$\pm$0.10 & 17.18$\pm$0.18 & 16.50$\pm$0.21 & 15.76$\pm$0.19 & ... & 17.25$\pm$0.02 & ... & ... & L2 \\
~~J083613.45$+$022106.2$^{\mathrm{c}}$ & 22.36$\pm$0.16 & 19.12$\pm$0.02 & 17.09$\pm$0.01 & 14.76$\pm$0.04 & 14.17$\pm$0.04 & 13.71$\pm$0.04 & 15.73$\pm$0.01 & 14.72$\pm$0.00 & 14.18$\pm$0.00 & 13.68$\pm$0.00 & L0.5 \\
J092745.81$+$010640.4 & 24.03$\pm$0.68 & 21.34$\pm$0.10 & 19.21$\pm$0.06 & 16.97$\pm$0.20 & 16.18$\pm$0.21 & 15.37$\pm$0.21 & 18.05$\pm$0.03 & 17.11$\pm$0.01 & 16.55$\pm$0.02 & 15.99$\pm$0.03 & L1 \\
J094624.37$+$344639.8 & 23.40$\pm$0.30 & 20.92$\pm$0.05 & 18.82$\pm$0.03 & 16.44$\pm$0.10 & 15.79$\pm$0.13 & 15.14$\pm$0.12 & ... & 16.33$\pm$0.01 & ... & ... & L1 \\
J095941.47$+$114146.0 & 23.51$\pm$0.38 & 21.30$\pm$0.08 & 19.29$\pm$0.05 & 16.43$\pm$0.14 & 15.66$\pm$0.13 & 15.24$\pm$0.19 & ... & ... & ... & 15.27$\pm$0.01 & L3 \\
J121238.73$+$000721.6 & 24.02$\pm$0.66 & 20.85$\pm$0.09 & 18.25$\pm$0.03 & 15.81$\pm$0.10 & 15.01$\pm$0.09 & 14.46$\pm$0.09 & ... & 15.69$\pm$0.01 & ... & ... & L3.5 \\
J133131.70$+$122531.4 & 24.62$\pm$0.66 & 21.32$\pm$0.10 & 19.32$\pm$0.05 & 16.78$\pm$0.16 & 15.72$\pm$0.17 & 15.28$\pm$0.15 & 17.83$\pm$0.03 & 16.72$\pm$0.01 & 16.03$\pm$0.02 & 15.46$\pm$0.01 & L2.5 \\
J134531.43$+$001551.2 & 24.14$\pm$0.51 & 21.59$\pm$0.11 & 19.47$\pm$0.08 & 16.94$\pm$0.21 & 15.98$\pm$0.14 & 15.55$\pm$0.23 & 18.19$\pm$0.03 & 16.95$\pm$0.02 & 16.19$\pm$0.02 & 15.60$\pm$0.02 & L2.5 \\
J150153.00$-$013507.1 & 22.71$\pm$0.25 & 20.80$\pm$0.07 & 18.62$\pm$0.05 & 16.08$\pm$0.09 & 14.95$\pm$0.07 & 14.26$\pm$0.09 & ... & ... & 15.02$\pm$0.01 & 14.25$\pm$0.01 & L3.5 \\
J154236.26$-$004545.9 & 24.71$\pm$0.79 & 21.78$\pm$0.15 & 19.35$\pm$0.06 & 16.71$\pm$0.13 & 15.98$\pm$0.14 & 15.41$\pm$0.20 & 18.07$\pm$0.02 & 16.83$\pm$0.02 & 16.21$\pm$0.02 & 15.64$\pm$0.02 & L3.5 \\
J154432.77$+$265551.2 & 23.22$\pm$0.23 & 21.31$\pm$0.07 & 19.10$\pm$0.04 & 16.22$\pm$0.10 & 15.24$\pm$0.10 & 14.64$\pm$0.10 & ... & 16.23$\pm$0.01 & ... & ... & L4.5 \\
J154740.16$+$053208.3 & 23.39$\pm$0.30 & 20.57$\pm$0.04 & 18.57$\pm$0.03 & 16.18$\pm$0.10 & 15.58$\pm$0.11 & 15.11$\pm$0.16 & 17.25$\pm$0.01 & 16.18$\pm$0.01 & 15.58$\pm$0.01 & 15.02$\pm$0.01 & L0.5 \\
J161711.68$+$322249.5 & 24.80$\pm$0.88 & 21.25$\pm$0.12 & 19.03$\pm$0.06 & 16.56$\pm$0.13 & 15.69$\pm$0.15 & 15.42$\pm$0.16 & ... & 16.68$\pm$0.01 & ... & ... & L2 \\
J232715.71$+$151730.4 & 23.26$\pm$0.28 & 21.31$\pm$0.09 & 19.13$\pm$0.05 & 16.29$\pm$0.11 & 15.35$\pm$0.09 & 14.77$\pm$0.13 & 17.54$\pm$0.02 & 16.20$\pm$0.01 & 15.36$\pm$0.01 & 14.68$\pm$0.01 & L5 \\
J234040.33$-$003337.2 & 23.54$\pm$0.35 & 21.45$\pm$0.10 & 19.35$\pm$0.06 & 17.06$\pm$0.16 & 16.34$\pm$0.23 & 15.52$\pm$0.20 & 17.94$\pm$0.03 & 17.03$\pm$0.02 & 16.46$\pm$0.02 & 16.01$\pm$0.04 & L1 \\
J234513.85$+$002441.6 & 24.80$\pm$0.71 & 22.63$\pm$0.29 & 19.56$\pm$0.08 & 16.78$\pm$0.16 & 15.90$\pm$0.20 & 15.55$\pm$0.19 & 17.67$\pm$0.03 & ... & 16.21$\pm$0.02 & 15.59$\pm$0.02 & L8.5 \\
\hline
\end{tabular}
\begin{list}{}{}
\item
Notes: UKIDSS magnitude limits for \emph{Y}, \emph{J}, \emph{H} and
\emph{K} bands are 20.5, 20.0, 18.8 and 18.4 respectively. All
magnitudes here are Vega based. We use 0.1$^{\prime\prime}$ as a
typical position error for UKIDSS. Error ellipses of 2MASS and SDSS
overlap for some objects for which proper motion angle errors are
not listed. \\
\item[$^{\mathrm{a}}$]Spectral types are based on the relationship between spectral
type and SDSS and 2MASS colors (\emph{i-z}, \emph{i-J}, \emph{i-H},
\emph{i-K} are used largely and
\emph{r-i}, \emph{z-J}, \emph{z-H}, \emph{z-K} with less weight). \\
\item[$^{\mathrm{b}}$]It is merged with a nearby faint
galaxy in 2MASS image, so its 2MASS photometric data is not
reliable. The spectral type is based on SDSS-UKIDSS colors
(\emph{i-J, i-H} and \emph{i-K}) according to equation (9).
\item[$^{\mathrm{c}}$]Data errors of \emph{J, H, K} bands are less
than 0.005.
\end{list}
\end{table}
\end{landscape}


\begin{table*}
 \caption{Seven objects with SDSS spectra matched in UKIDSS} \centering \label{tab3}
 \begin{tabular}{c c c c c c l} \hline\hline
SDSS Name & UKIDSS \emph{Y} & UKIDSS \emph{J} & UKIDSS \emph{H} & UKIDSS \emph{K} & Proper Motion$^{\mathrm{a}}$ & Proper Motion \\
 & & & & & ($^{\prime\prime}$yr$^{-1}$) & Angle$^{\mathrm{b}}$ \\
\hline
SDSS J012052.58+151827.3 & 17.11$\pm$0.02  & 16.09$\pm$0.01  & 15.50$\pm$0.01  & 14.92$\pm$0.01  &  0.04$\pm$0.02 & 105$\pm$36 \\
SDSS J084751.48+013811.0 & 17.38$\pm$0.02  & 16.09$\pm$0.01  & 15.22$\pm$0.01  & 14.45$\pm$0.01  &  0.03$\pm$0.02 & 242 ... \\
SDSS J090347.55+011446.0 & 17.58$\pm$0.02  & ... & 15.59$\pm$0.01  & 14.96$\pm$0.01  &   0.03$\pm$0.02 & 259 ... \\
SDSS J091714.76+314824.8 & ... & 16.36$\pm$0.01  & ... & ... & 0.03$\pm$0.03 & 345 ... \\
SDSS J100817.07+052312.9 & ... & ... & 16.36$\pm$0.02  & 15.89$\pm$0.03  &  0.21$\pm$0.03 & 139$\pm$10 \\
SDSS J154502.87+061807.8 & 17.35$\pm$0.02  & 16.21$\pm$0.01  & 15.62$\pm$0.01  & 15.07$\pm$0.01  &   0.14$\pm$0.04 & 251$\pm$25 \\
SDSS J155215.38+065041.5 & ... & ... & 15.93$\pm$0.01  & 15.32$\pm$0.01  & 0.02$\pm$0.04 & 356 ... \\
\hline
\end{tabular}
\begin{list}{}{}
\item[$^{\mathrm{a}}$]SDSS-UKIDSS data-base proper motions - found by dividing the difference between
the SDSS and UKIDSS coordinates (from the respective databases) by
the observational epoch difference.
Standard errors are calculated using the major axes of the position error ellipses from SDSS and UKIDSS.\\
\item[$^{\mathrm{b}}$]Error ellipses of SDSS and UKIDSS overlap for some
objects for which position angle errors are not meaningful.\\
\end{list}
\end{table*}

Using the relationship between absolute \emph{J} and \emph{z} band
magnitudes and spectral types (Hawley et al. \cite{haw02}), we
estimated the approximate distance of the 36 new M and L dwarfs.
Generally, early and mid L dwarfs in our sample are between 25 and
100 pc, and the M dwarfs beyond 100 pc.



\begin{table*}
 \caption{Parameters of
fitting equations for color$-$type relationships.} \centering
\label{tab4}
\begin{tabular}{crrrccc}
\hline\hline
Color & \emph{a} & \emph{b} & \emph{c} & Type & R & Sensitivity \\
 &  &  &  & Range & & Index$^{\mathrm{a}}$    \\
\hline
r$-$i & 2.77242  & 0.01930 & $-$0.00355 & [10,27.5] & 0.62  & 0.073 \\
i$-$z & 0.89637  & 0.00812 & 0.00582  & [13,25] & 0.89  & 0.225 \\
i$-$J & 4.46097  & $-$0.12872 & 0.01099  & [10,25] & 0.94  & 0.253 \\
i$-$H & 2.68743  & 0.22692 & 0 & [10,25] & 0.92  & 0.227 \\
i$-$K & 2.04847  & 0.39856 & $-$0.00567 & [10,25] & 0.90  & 0.200 \\
z$-$J & 1.32530  & 0.10725 & 0 & [10,13.5] & 0.59  & 0.107 \\
z$-$J & 0.61620  & 0.10826 & 0 & [19,26] & 0.81  & 0.107 \\
z$-$H & 4.23705  & $-$0.31004 & 0.02038  & [10,14] & 0.70  & 0.158 \\
z$-$H & 4.60067  & $-$0.03962 & 0 & [17.5,27.5] & 0.53  & 0.040 \\
z$-$K & 3.93079  & $-$0.21325 & 0.01860  & [10,14] & 0.72  & 0.235 \\
z$-$K & 6.96601  & $-$0.13083 & 0 & [17.5,27.5] & 0.72  & 0.080 \\
H$-$K & 0.15050  & 0.03638 & 0 & [10,15] & 0.40  & 0.040 \\
H$-$K & 2.19557  & $-$0.08809 & 0 & [17,25] & 0.66  & 0.088 \\
J$-$H & 0.06224  & 0.06903 & 0 & [10,14] & 0.49  & 0.088 \\
J$-$H & 5.58544  & $-$0.21707 & 0 & [20,25] & 0.90  & 0.100 \\
\hline
\end{tabular}
\begin{list}{}{}
\item
Notes: The united polynomial fitting equation of color$-$spectral
type relationships is: $color=a+b(type)+c(type)^{2}$, type$=$10 for
L0, 15 for L5, 20
for T0, 25 for T5. \\
\item[$^{\mathrm{a}}$]Sensitivity indices are the rates
of change of color ranges with spectral ranges covered by fitting
lines.
\end{list}
\end{table*}

\section{Color-spectral type relationships}
\begin{figure}
   \centering
   \includegraphics[width=9cm]{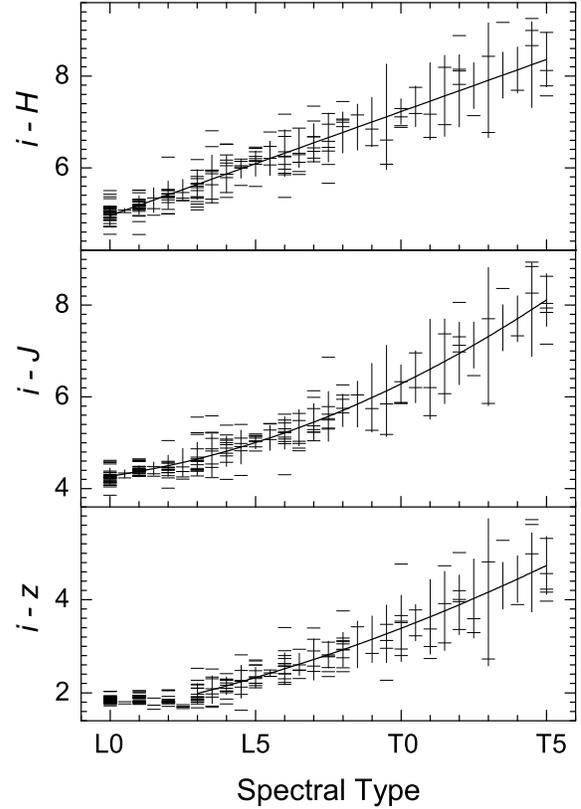}
      \caption{Polynomial fitting for
    color-spectral type relationships are indicated with a solid line.
    The error bars of fitted colors indicate the standard deviation at
    each subtype. We used 0.5 as the error, if there is only one object
    available for a given subtype.}
         \label{f6}
   \end{figure}

To estimate the spectral types of our ultra-cool dwarf candidates
without spectra, we need to know the relationships between spectral
types and colors. Hawley et al. (\cite{haw02}) gave the correlation
between spectral type and average color for each subtype range from
M0 to T6. The relationship is good for M dwarfs, but has large
errors for L and T dwarfs. With a much larger number of L and T
dwarfs now available, we made a study of the relationships between
spectral types of L and T dwarfs and their colors from SDSS and
2MASS. To construct these relationships we used the same data set as
that for the photometric selection criteria (see section 2). SDSS
colors \emph{r-i, i-z}, 2MASS colors \emph{J-H}, \emph{H-K} and
optical-near infrared colors \emph{i-J}, \emph{i-H}, \emph{i-K}, \emph{z-J},
\emph{z-H}, \emph{z-K} are involved. We fit these relationships with
a united polynomial equation,
\begin{equation}
color=a+b(type)+c(type)^{2}
\end{equation}
where type is a number designed to incompass the full range of M, L
and T spectral classes (type$=$10 for L0, 15 for L5, 20 for T0, 25
for T5). Polynomial parameters \emph{a}, \emph{b} and \emph{c} are
different for different color ranges and our range of calculated
values can be found in Table 4. As well as spectral type ranges,
correlation coefficient R and sensitivity indices for the fitting
equations are also available in Table 4, where the sensitivity index
is defined as the rate of change of color with spectral type, and is
thus an indication of the usefulness of a color as a spectral type
estimator. Calculated values of the various colors are presented in
Table 5 for spectral sub-types between L0 and T7.5. The \emph{i-z},
\emph{i-J}, \emph{i-H} and \emph{i-K} colors are the most sensitive
to spectral type, and the first three of these are plotted as a
function of spectral type in Figure~\ref{f6}.

\section{Cross matching the new sample with UKIDSS}
To provide an additional epoch of deeper near infrared measurements
that could improve candidate characterization (particularly for
candidates without SDSS spectroscopy), we cross-matched our SDSS DR7
candidates with the Fourth Data Release of the UKIDSS Large Area
Survey (Lawrence et al. \cite{law07}). UKIDSS magnitude limits for
the \emph{Y}, \emph{J}, \emph{H} and \emph{K} bands are 20.5, 20.0,
18.8 and 18.4 respectively. We found that 23 of our candidates
(without SDSS spectra) and 7 of our spectroscopically confirmed
objects have UKIDSS DR4 counterparts. Table 2 lists the SDSS names,
\emph{r}, \emph{i}, \emph{z}, 2MASS \emph{J}, \emph{H}, \emph{K},
UKIDSS \emph{Y}, \emph{J}, \emph{H}, \emph{K} and color-estimated
spectral types (see Section 4) for the objects without spectroscopy,
and Table 3 presents the additional UKIDSS information for 7 of the
spectroscopic objects from Table 1.

\begin{figure}
\centering
\includegraphics[width=9.5cm]{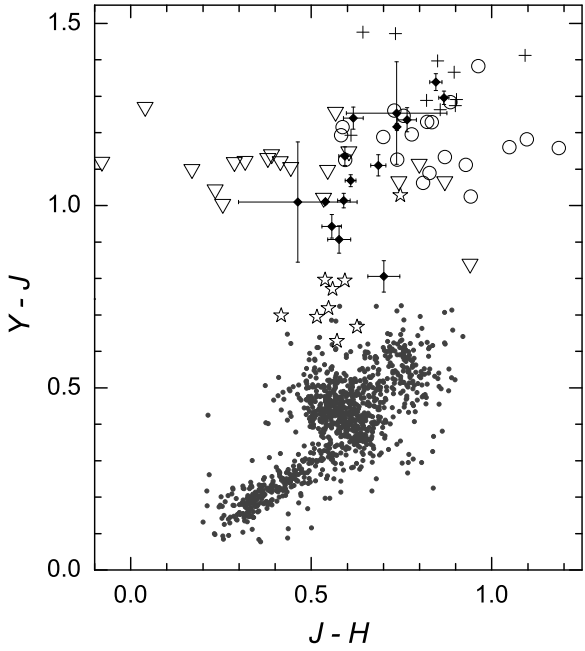}
\caption{$Y-J$ vs. $J-H$ diagram for known M, L and T dwarfs and 15
ultracool dwarf candidates (\emph{diamonds}) which have \emph{Y}
band photometric data from UKIDSS, include 4 objects with SDSS
spectra. M5.5-M8.5 (\emph{open pentacles}), L0-L4.5
(\emph{crosses}), L5-L9.5 (\emph{open circles}), T0-T3.5 (\emph{open
triangles}) and 1024 sources from UKIDSS LAS in 1 $deg^2$ with $Y <
18.5$ (\emph{gray points}). The \emph{J} band photometric data of
the two with the very large errors used here are transformed from
2MASS.} \label{f7}
\end{figure}

Figure~\ref{f7} shows the $Y-J$ versus $J-H$ diagram for the 15
candidates that had UKIDSS $Y$ detections. In general these all had
UKIDSS $YJH$ magnitudes but in 2 cases we had to transform a 2MASS
\emph{J} into a UKIDSS $J$ following the conversion of Hewett et al.
\cite{hew06}. This 2-color diagram can provide additional
information on ultra-cool dwarf spectral types (e.g. Hewett et al.
\cite{hew06}). For example, the SDSS/2MASS colors of SDSS
J2345$-$0024 suggest a spectral type between L7 and T2. However,
Figure 7 suggests that SDSS J2345$-$0024 is more likely an early T
dwarf than a late L dwarf.

\begin{figure}
\centering
\includegraphics[width=9.5cm]{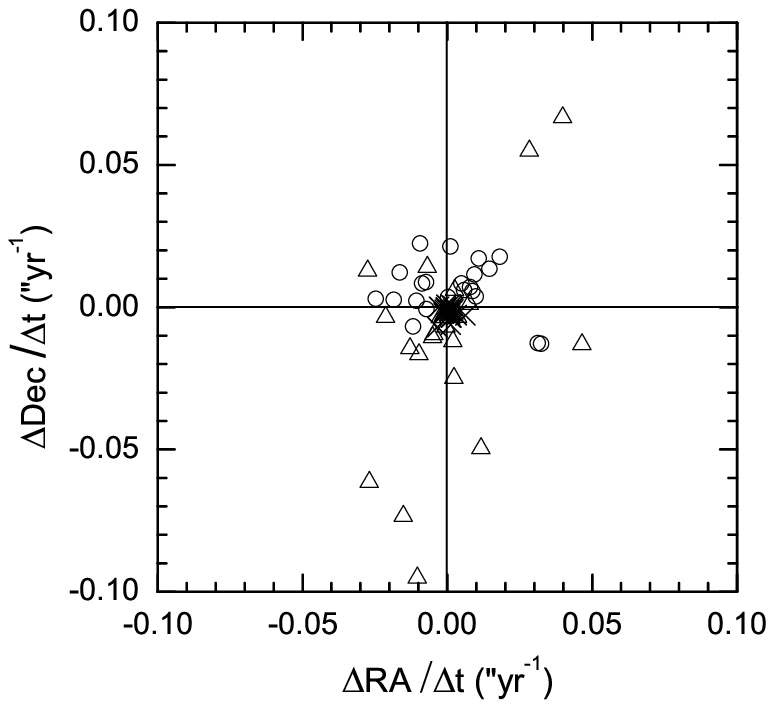}
\caption{Average proper motions of reference objects for ultracool
dwarf candidates in Table 6. Different symbols indicated systematic
coorections used for different survey combinations, 2MASS-SDSS
(\emph{open triangles}, 21 objects), SDSS-UKIDSS (\emph{open
circles}, 23 objects) and 2MASS-UKIDSS (\emph{crosses}, 23
objects).} \label{f8}
\end{figure}

To further assess the proper motions that we derive for all
candidates using the coordinate and epoch information stored in the
SDSS and 2MASS databases, we have measured additional proper motion
constraints using several combinations of multi-epoch data
(SDSS/UKIDSS, 2MASS/UKIDSS and 2MASS/SDSS combinations), as well as
by measuring the motion of our ultracool dwarf candidates with
respect to reference objects within 6$^{\prime}$ in the SDSS
\emph{z} and UKIDSS \emph{Y, J, H} and \emph{K} images. For the
relative proper motions, we used the Iraf routines GEOMAP and
GEOXYTRAN to transform the pixel coordinates from the SDSS images
into the pixel coordinate system of the UKIDSS images. Visual
inspection of the image data revealed a small number of problematic
sources. SDSS J0047$+$1353 is merged with a very nearby galaxy in
the 2MASS image. In the main however, four separate proper motion
measurements were made (combining $z/Y, z/J, z/H$ and $z/K$) where
possible, and an average proper motion taken. UKIDSS centroiding
accuracy was estimated from the standard deviation of these four
measurements, and we also factored in a centroiding uncertainty
associated with the SDSS $z$-band epoch, which we estimated to be
0.1$^{\prime\prime}$ (1/4 of an SDSS pixel) for these faint sources.
The relative proper motions measured from 2MASS and SDSS images are
given in columns 10 and 11 of Table 1 (for the spectroscopically
confirmed objects). The relative proper motions measured from UKIDSS
and SDSS images are given in columns 2 and 3 of Table 6. Database
proper motions (i.e. calculated from cross-database coordinate/epoch
information) for UKIDSS/SDSS, 2MASS/UKIDSS, and 2MASS/SDSS database
combinations are given in columns 4-9 of Table 6, and for the
UKIDSS/SDSS combination in columns 6-7 of Table 3. A correction for
systematic coordinate uncertainties between UKIDSS and 2MASS/SDSS
was decreased as before (see section 3). Figure 8 shows average
proper motions of reference objects of objects in Table 6 which
proper motions measured with 2MASS-SDSS-UKIDSS databases. The proper
motion offsets of 2MASS-UKIDSS with the longest baseline have the
smallest offsets ($<$0.008$^{\prime\prime}$yr$^{-1}$) which
indicated proper motion with a baseline of longer than 5 years (e.g.
2MASS-UKIDSS) will be a very good way for identified ultracool
dwarfs.

\begin{itemize}

\item{In most cases the 2MASS-SDSS proper motions calculated from the databases
are in reasonable agreement with our derived relative proper motions
to within the uncertainties. However, the uncertainties associated
with the relative proper motions are often larger. This is because
in general there is a reduced number of useful reference sources in
the 2MASS images. This can sometimes be compounded if a sources is
close to the edge of a 2MASS strip, since the number of reference
sources can be reduced still further. We thus conclude that the
database proper motions are to be preferred when combining 2MASS and
SDSS survey data.}

\item{In general, the UKIDSS/SDSS, 2MASS/UKIDSS and 2MASS/SDSS database proper
motions agree well to within their uncertainties, except in a limited number
of cases, where we find that on closer inspection the baseline between the
database epochs is low and the 2MASS sources themselves are near the 2MASS
detection limit. The 2MASS/SDSS database proper motions thus stand up reasonably
well when compared to those measured from a combination of, on average, higher
signal-to-noise imaging data.}

\item{The UKIDSS-SDSS proper motions calculated from the databases are in very good
agreement with our derived relative proper motions to within the
uncertainties, and the estimated uncertainties from each method are
comparable. This result is not surprising, but provides a useful
verification for our database proper motion calculations.}

\end{itemize}

We thus conclude that overall, our 2MASS-SDSS proper motions calculated from
the databases offer a good balance of reasonably accurate measurements over
a relatively large sky area.


\begin{landscape}
\begin{table}
\caption{Different colors by different spectral types.} \centering
\label{tab5}
\begin{tabular}{lcccccccccc}
\hline\hline
Sp.Type & \emph{r-i} & \emph{i-z} & \emph{i-J} & \emph{i-H} & \emph{i-K} & \emph{z-J}$^{a}$ & \emph{z-H}$^{a}$ & \emph{z-K$^{a}$} & \emph{J-H}$^{a}$ & \emph{H-K}$^{a}$ \\
\hline
L0 & 2.61$\pm$0.27 & ... & 4.27$\pm$0.17 & 4.96$\pm$0.22 & 5.47$\pm$0.26 & 2.40$\pm$0.13 & 3.17$\pm$0.17 & 3.66$\pm$0.22 & 0.75$\pm$0.11 & 0.51$\pm$0.10 \\
L0.5 & 2.58$\pm$0.35 & ... & 4.32$\pm$0.09 & 5.07$\pm$0.05 & 5.61$\pm$0.10 & 2.45$\pm$0.13 & 3.23$\pm$0.09 & 3.74$\pm$0.21 & 0.79$\pm$0.09 & 0.53$\pm$0.07 \\
L1 & 2.56$\pm$0.19 & ... & 4.37$\pm$0.12 & 5.18$\pm$0.20 & 5.75$\pm$0.21 & 2.51$\pm$0.12 & 3.29$\pm$0.22 & 3.84$\pm$0.23 & 0.82$\pm$0.14 & 0.55$\pm$0.12 \\
L1.5 & 2.52$\pm$0.10 & ... & 4.43$\pm$0.14 & 5.30$\pm$0.27 & 5.88$\pm$0.13 & 2.56$\pm$0.06 & 3.37$\pm$0.13 & 3.94$\pm$0.13 & 0.86$\pm$0.14 & 0.57$\pm$0.09 \\
L2 & 2.49$\pm$0.28 & ... & 4.50$\pm$0.23 & 5.41$\pm$0.30 & 6.01$\pm$0.38 & 2.61$\pm$0.21 & 3.45$\pm$0.29 & 4.05$\pm$0.39 & 0.89$\pm$0.18 & 0.59$\pm$0.14 \\
L2.5 & 2.46$\pm$0.33 & ... & 4.57$\pm$0.31 & 5.52$\pm$0.23 & 6.14$\pm$0.29 & 2.67$\pm$0.14 & 3.55$\pm$0.16 & 4.17$\pm$0.23 & 0.93$\pm$0.14 & 0.61$\pm$0.13 \\
L3 & 2.42$\pm$0.32 & 1.99$\pm$0.21 & 4.64$\pm$0.36 & 5.64$\pm$0.32 & 6.27$\pm$0.43 & 2.72$\pm$0.19 & 3.65$\pm$0.26 & 4.30$\pm$0.33 & 0.96$\pm$0.17 & 0.62$\pm$0.17 \\
L3.5 & 2.39$\pm$0.64 & 2.07$\pm$0.30 & 4.73$\pm$0.47 & 5.75$\pm$0.53 & 6.40$\pm$0.57 & 2.77$\pm$0.38 & 3.77$\pm$0.48 & 4.44$\pm$0.48 & 0.99$\pm$0.23 & 0.64$\pm$0.24 \\
L4 & 2.35$\pm$0.56 & 2.15$\pm$0.19 & 4.81$\pm$0.33 & 5.86$\pm$0.40 & 6.52$\pm$0.55 & ... & 3.89$\pm$0.30 & 4.59$\pm$0.46 & 1.03$\pm$0.28 & 0.66$\pm$0.17 \\
L4.5 & 2.31$\pm$0.55 & 2.24$\pm$0.35 & 4.91$\pm$0.37 & 5.98$\pm$0.21 & 6.64$\pm$0.30 & ... & ... & ... & ... & 0.68$\pm$0.21 \\
L5 & 2.26$\pm$0.53 & 2.33$\pm$0.22 & 5.00$\pm$0.14 & 6.09$\pm$0.27 & 6.75$\pm$0.32 & ... & ... & ... & ... & 0.70$\pm$0.13 \\
L5.5 & 2.22$\pm$0.38 & 2.42$\pm$0.07 & 5.11$\pm$0.27 & 6.20$\pm$0.37 & 6.86$\pm$0.45 & ... & ... & ... & ... & ... \\
L6 & 2.17$\pm$0.75 & 2.52$\pm$0.34 & 5.21$\pm$0.35 & 6.32$\pm$0.51 & 6.97$\pm$0.51 & ... & ... & ... & ... & ... \\
L6.5 & 2.12$\pm$0.55 & 2.61$\pm$0.24 & 5.33$\pm$0.39 & 6.43$\pm$0.43 & 7.08$\pm$0.38 & ... & ... & ... & ... & ... \\
L7 & 2.07$\pm$0.54 & 2.72$\pm$0.44 & 5.45$\pm$0.40 & 6.55$\pm$0.42 & 7.19$\pm$0.48 & ... & ... & ... & ... & 0.70$\pm$0.27 \\
L7.5 & 2.02$\pm$0.93 & 2.82$\pm$0.29 & 5.57$\pm$0.54 & 6.66$\pm$0.53 & 7.29$\pm$0.55 & ... & 3.91$\pm$0.42 & 4.68$\pm$0.53 & ... & 0.65$\pm$0.18 \\
L8 & 1.97$\pm$0.63 & 2.93$\pm$0.37 & 5.70$\pm$0.35 & 6.77$\pm$0.45 & 7.39$\pm$0.58 & ... & 3.89$\pm$0.17 & 4.61$\pm$0.21 & ... & 0.61$\pm$0.13 \\
L8.5$^{\mathrm{b}}$ & 1.91 ~~...~~~ & 3.04 ~~...~~~ & 5.84 ~~...~~~ & 6.89 ~~...~~~ & 7.48 ~~...~~~ & ... & 3.87 ~~...~~~ & 4.55 ~~...~~~ & ... & 0.57 ~~...~~~ \\
L9$^{\mathrm{b}}$ & 1.86 ~~...~~~ & 3.15 ~~...~~~ & 5.98$\pm$0.75 & 7.00$\pm$0.53 & 7.57$\pm$0.49 & 2.67$\pm$0.27 & 3.85$\pm$0.28 & 4.48$\pm$0.27 & ... & 0.52$\pm$0.15 \\
L9.5 & 1.80$\pm$1.38 & 3.27$\pm$0.62 & 6.13$\pm$0.99 & 7.11$\pm$1.15 & 7.66$\pm$1.10 & 2.73$\pm$0.18 & 3.83$\pm$0.34 & 4.41$\pm$0.24 & ... & 0.48$\pm$0.20 \\
T0 & 1.74$\pm$1.49 & 3.39$\pm$0.71 & 6.28$\pm$0.42 & 7.23$\pm$0.28 & 7.75$\pm$0.31 & 2.78$\pm$0.29 & 3.81$\pm$0.22 & 4.35$\pm$0.15 & 1.24$\pm$0.23 & 0.43$\pm$0.14 \\
T0.5 & 1.68$\pm$0.65 & 3.51$\pm$0.39 & 6.44$\pm$0.57 & 7.34$\pm$0.44 & 7.84$\pm$0.53 & 2.84$\pm$0.37 & 3.79$\pm$0.25 & 4.28$\pm$0.55 & 1.14$\pm$0.37 & 0.39$\pm$0.45 \\
T1 & 1.61$\pm$0.65 & 3.63$\pm$0.80 & 6.60$\pm$1.09 & 7.45$\pm$0.84 & 7.92$\pm$0.71 & 2.89$\pm$0.30 & 3.77$\pm$0.10 & 4.22$\pm$0.08 & 1.03$\pm$0.09 & 0.35$\pm$0.20 \\
T1.5 & 1.55$\pm$0.42 & 3.76$\pm$0.85 & 6.77$\pm$0.92 & 7.57$\pm$0.88 & 8.00$\pm$0.72 & 2.94$\pm$0.31 & 3.75$\pm$0.31 & 4.15$\pm$0.66 & 0.92$\pm$0.13 & 0.30$\pm$0.24 \\
T2 & 1.48$\pm$1.02 & 3.89$\pm$0.64 & 6.95$\pm$0.68 & 7.68$\pm$0.79 & 8.07$\pm$0.71 & 3.00$\pm$0.10 & 3.73$\pm$0.28 & 4.09$\pm$0.33 & 0.81$\pm$0.17 & 0.26$\pm$0.53 \\
T2.5$^{\mathrm{b}}$ & 1.41$\pm$0.83 & 4.03$\pm$0.85 & 7.13 ~~...~~~ & 7.79 ~~...~~~ & 8.15 ~~...~~~ & 3.05 ~~...~~~ & 3.71 ~~...~~~ & 4.02 ~~...~~~ & 0.70 ~~...~~~ & 0.21 ~~...~~~ \\
T3 & 1.34$\pm$2.19 & 4.16$\pm$1.58 & 7.31$\pm$1.51 & 7.91$\pm$1.25 & 8.22$\pm$1.19 & 3.11$\pm$0.22 & 3.69$\pm$0.36 & 3.96$\pm$0.50 & 0.59$\pm$0.25 & 0.17$\pm$0.16 \\
T3.5$^{\mathrm{b}}$ & 1.27 ~~...~~~ & 4.30 ~~...~~~ & 7.51 ~~...~~~ & 8.02 ~~...~~~ & 8.28 ~~...~~~ & 3.16 ~~...~~~ & 3.67 ~~...~~~ & 3.89 ~~...~~~ & 0.48 ~~...~~~ & 0.13 ~~...~~~ \\
T4$^{\mathrm{b}}$ & 1.19 ~~...~~~ & 4.44 ~~...~~~ & 7.70 ~~...~~~ & 8.13 ~~...~~~ & 8.35 ~~...~~~ & 3.21 ~~...~~~ & 3.65 ~~...~~~ & 3.83 ~~...~~~ & 0.38 ~~...~~~ & 0.08 ~~...~~~ \\
T4.5 & 1.11$\pm$1.24 & 4.59$\pm$0.85 & 7.90$\pm$1.02 & 8.25$\pm$0.93 & 8.41$\pm$1.11 & 3.27$\pm$0.11 & 3.63$\pm$0.26 & 3.76$\pm$0.81 & 0.27$\pm$0.20 & 0.04$\pm$0.51 \\
T5 & 1.04$\pm$0.83 & 4.74$\pm$0.62 & 8.11$\pm$0.57 & 8.36$\pm$0.60 & 8.47$\pm$0.79 & 3.32$\pm$0.21 & 3.61$\pm$0.10 & 3.70$\pm$0.40 & 0.16$\pm$0.19 & $-$0.01$\pm$0.27 \\
T5.5 & 0.96$\pm$1.45 & ... & ... & ... & ... & 3.38$\pm$0.04 & 3.59$\pm$0.16 & 3.63$\pm$0.73 & ... & ... \\
T6 & 0.87$\pm$1.79 & ... & ... & ... & ... & 3.43$\pm$0.26 & 3.57$\pm$0.36 & 3.56$\pm$0.83 & ... & ... \\
T6.5$^{\mathrm{b}}$ & 0.79 ~~...~~~ & ... & ... & ... & ... & ... & 3.55$\pm$0.20 & 3.50$\pm$0.50 & ... & ... \\
T7 & 0.71$\pm$0.77 & ... & ... & ... & ... & ... & 3.53$^{\mathrm{c}}$ ~...~~~ & 3.43$^{\mathrm{c}}$ ~...~~~ & ... & ... \\
T7.5$^{\mathrm{b}}$ & 0.62 ~~...~~~ & ... & ... & ... & ... & ... & 3.51 ~~...~~~ & 3.37 ~~...~~~ & ... & ... \\
\hline
\end{tabular}
\begin{list}{}{}
\item
Notes: Colors of different spectral types are calculated with our
fitting equation (equation 9). Standard errors are used. \\
\item[$^{\mathrm{a}}$]For these colors, the spectral type-color
relationships can be double valued. One can relay or indications
from the other colors (\emph{i-J, i-H} or \emph{i-z}) to identify
the correct value to use. \\
\item[$^{\mathrm{b}}$]There is only one object available for
colors of these subtypes without errors, we prefer to use 0.5 as
their errors. \\
\item[$^{\mathrm{c}}$]No object with this type has data of
these colors.
\end{list}
\end{table}
\end{landscape}


\begin{landscape}
\begin{table}
 \caption{Proper motions constraints of SDSS L dwarfs that were also found in UKIDSS DR4.}
 \centering \label{tab6}
\begin{tabular}{cllclclcl}
\hline\hline
SDSS Name & Proper Motion$^{\mathrm{a}}$ & Proper Motion & Proper Motion$^{\mathrm{b}}$ & Proper Motion & Proper Motion$^{\mathrm{c}}$ & Proper Motion & Proper Motion$^{\mathrm{d}}$ & Proper Motion \\
 & ($^{\prime\prime}$yr$^{-1}$) & Angle$^{\mathrm{a}}$ & ($^{\prime\prime}$yr$^{-1}$) & Angle$^{\mathrm{e}}$ & ($^{\prime\prime}$yr$^{-1}$) & Angle$^{\mathrm{e}}$ & ($^{\prime\prime}$yr$^{-1}$) & Angle$^{\mathrm{e}}$       \\
\hline
SDSS J004759.59$+$135332.0 & ~~~~~~~~...$^{\mathrm{h}}$ & ~~~~~~~~... &0.13$\pm$0.02 &208$\pm$10&0.13$\pm$0.03 &234$\pm$14&0.47$\pm$0.27 &300$\pm$36 \\
SDSS J015141.04$-$005156.5 & 0.044$\pm$0.012  & 274.8$\pm$14.2  &0.07$\pm$0.06 &285$\pm$57&0.04$\pm$0.03 &257$\pm$72&0.07$\pm$0.15 &150 ... \\
SDSS J022927.95$-$005328.5 & 0.068$\pm$0.012  & 104.8$\pm$25.1  &0.02$\pm$0.03 &~~57 ...&0.04$\pm$0.04 &136 ...&0.10$\pm$0.09 &143$\pm$63 \\
SDSS J073241.77$+$264558.9 & 0.002 ... & ~~30.3 ... &0.04$\pm$0.02 &359$\pm$32&0.02$\pm$0.03 &230 ...&0.12$\pm$0.06 &206$\pm$30 \\
SDSS J074436.02$+$251330.5 & 0.033 ... & 262.6 ... &0.04$\pm$0.03 &296$\pm$50&0.06$\pm$0.06 &296 ...&0.05$\pm$0.28 &322 ... \\
SDSS J075754.16$+$221604.9 & 0.016 ... & ~~74.6 ... &0.02$\pm$0.04 &~~63 ...&0.05$\pm$0.04 &116$\pm$59&0.11$\pm$0.07 &124$\pm$38 \\
SDSS J081303.96$+$243355.9 & 0.066 ... & 251.3 ... &0.06$\pm$0.03 &254$\pm$30&0.08$\pm$0.03 &228$\pm$25&0.09$\pm$0.06 &216$\pm$42 \\
SDSS J081409.45$+$260250.4 & 0.057 ... & 278.5 ... &0.06$\pm$0.03 &304$\pm$35&0.03$\pm$0.03 &217$\pm$59&0.08$\pm$0.05 &170$\pm$36 \\
SDSS J083613.45$+$022106.2 & 0.096$\pm$0.003  & 285.0$\pm$1.6  &0.07$\pm$0.02 &311$\pm$17&0.11$\pm$0.03 &291$\pm$14&0.36$\pm$0.16 &263$\pm$26 \\
SDSS J092745.81$+$010640.4 & 0.100$\pm$0.001  & 274.1$\pm$2.0  &0.12$\pm$0.02 &272$\pm$~~9&0.10$\pm$0.03 &302$\pm$19& ...$^{\mathrm{g}}$ &~~37$\pm$58 \\
SDSS J094624.37$+$344639.8 & 0.083 ... & 293.8 ... &0.09$\pm$0.03 &304$\pm$22&0.07$\pm$0.04 &298$\pm$33&0.05$\pm$0.07 &295 ... \\
SDSS J095941.47$+$114146.0 & 0.148 ... & 254.1 ... &0.15$\pm$0.03 &259$\pm$12&0.14$\pm$0.04 &257$\pm$18&0.11$\pm$0.10 &252$\pm$67 \\
SDSS J121238.73$+$000721.6 & 0.037 ... & 278.5 ... &0.03$\pm$0.02 &342$\pm$38&0.04$\pm$0.05 &282 ... & ~0.13$\pm$0.27$^{\mathrm{f}}$  &~~20 ... \\
SDSS J133131.70$+$122531.4 & 0.182$\pm$0.006  & 193.7$\pm$1.4  &0.16$\pm$0.03 &188$\pm$11&0.12$\pm$0.02 &207$\pm$~~8& 0.12$\pm$0.03  &243$\pm$14 \\
SDSS J134531.43$+$001551.2 & 0.047$\pm$0.006  & ~~95.2$\pm$1.9  &0.05$\pm$0.02 &~~77$\pm$22&0.04$\pm$0.05 &124 ...& ~0.25$\pm$0.27$^{\mathrm{f}}$  &~~34 ... \\
SDSS J150153.00$-$013507.1 & 0.236$\pm$0.002  & 255.3$\pm$0.7  &0.26$\pm$0.03 &264$\pm$~~6&0.24$\pm$0.02 &258$\pm$~~5&0.21$\pm$0.07 &241$\pm$19 \\
SDSS J154236.26$-$004545.9 & 0.531$\pm$0.001  & 258.5$\pm$0.1  &0.52$\pm$0.02 &259$\pm$~~2&0.60$\pm$0.05 &252$\pm$~~5& ...$^{\mathrm{g}}$ &~~40$\pm$25 \\
SDSS J154432.77$+$265551.2 & 0.141 ... & 311.9 ... &0.16$\pm$0.04 &310$\pm$15&0.15$\pm$0.03 &304$\pm$11&0.18$\pm$0.05 &295$\pm$16 \\
SDSS J154740.16$+$053208.3 & 0.047$\pm$0.006  & 293.1$\pm$3.9  &0.06$\pm$0.05 &327$\pm$46&0.06$\pm$0.05 &243$\pm$54&0.16$\pm$0.11 &209$\pm$43 \\
SDSS J161711.68$+$322249.5 & 0.099 ... & 213.0 ... &0.04$\pm$0.04 &~~61 ...&0.02$\pm$0.04 &107 ...&0.02$\pm$0.06 &173 ... \\
SDSS J232715.71$+$151730.4 & 0.160$\pm$0.001  & 202.4$\pm$0.7 &0.20$\pm$0.02 &216$\pm$~~5&0.18$\pm$0.02 &211$\pm$~~5&0.10$\pm$0.04 &210$\pm$28 \\
SDSS J234040.33$-$003337.2 & 0.028$\pm$0.003  & 258.8$\pm$6.4  &0.05$\pm$0.04 &~~49$\pm$57&0.01$\pm$0.04 &199 ...&0.05$\pm$0.08 &214 ... \\
SDSS J234513.85$+$002441.6 & 0.092$\pm$0.001  & 200.1$\pm$4.2  &0.29$\pm$0.03&189$\pm$~~7&0.05$\pm$0.05 &221 ...&1.04$\pm$0.26 &~~~~4$\pm$14 \\
\hline
\end{tabular}
\begin{list}{}{}
\item[$^{\mathrm{a}}$]SDSS-UKIDSS relative proper motions - found by specifically measuring the relative
movement of the ultracool dwarfs with respect to nearby reference objects in the SDSS and UKIDSS images. \\
\item[$^{\mathrm{b}}$]SDSS-UKIDSS data-base proper motions - found by dividing the difference between
the SDSS and UKIDSS coordinates (from the respective databases) by
the observational epoch difference.
Standard errors are calculated using the major axes of the position error ellipses from SDSS and UKIDSS.\\
\item[$^{\mathrm{c}}$]2MASS-UKIDSS database proper motions - as for $b$ except using 2MASS and UKIDSS.\\
\item[$^{\mathrm{d}}$]2MASS-SDSS database proper motions - as for $b$ except using 2MASS and SDSS. We do not present
the proper motions calculated with a baseline 10 months. \\
\item[$^{\mathrm{e}}$]Error ellipses of 2MASS and SDSS overlap for some objects for which position angle
errors are not meaningful.\\
\item[$^{\mathrm{f}}$]Objects with a 2MASS-SDSS baseline $<$ 1 year.
\item[$^{\mathrm{g}}$]We do not present their proper motions for they have a short 2MASS-SDSS baseline ($<$ 2 months).
\item[$^{\mathrm{h}}$]We do not measure its proper motion for it is very faint in SDSS \emph{z}-band
image.
\end{list}
\end{table}
\end{landscape}

\section{Discussion}

Although our photometric selection criteria have been shown to be
optimized for mid-late L dwarfs (see Figure 1), most of the sample
that had SDSS spectra are actually late M and early L dwarfs. This
partly results from a luminosity bias since later, less luminous L
dwarfs are only detected by SDSS in a smaller volume. SDSS targeting
priority for these objects was primarily determined through
brightness, so our spectroscopic sample is reasonably close to a
magnitude limited subset of our full photometric selection. However,
in addition, later L dwarfs are redder, and are more likely to
become \emph{i}-band drop outs, precluding their selection in our
sample. For our full sample, we reach fainter magnitudes and thus
identify more later L dwarfs. The spectral type distribution of the
candidates without spectra (i.e. based on the relationship between
spectral type and colors) spans a range out to T3, with many
candidates in the L0-L7 range.

Our spectral typing procedure makes use of numerous optical, near
infrared and optical-infrared colors. Overall we find that the
$i-z$, $i-J$, $i-H$ and $i-K$ colors are the most useful. For the
objects in Table 1, the spectral types based on SDSS spectra and
those based on colors generally agree with each other well. However,
we caution against the use of $i-z$ for estimating spectral types
earlier than L3, as Figure~\ref{f3} shows \emph{i-z} does not
correlate well in this range. In our analysis, if the \emph{i-J} and
\emph{i-H} colors indicate a spectral type earlier than L3, then we
do not include an estimated type based on the \emph{i-z} color. The
addition of UKIDSS photometry adds an additional means to constrain
spectral type (e.g. using $Y-J$), particularly in and around the L-T
transition ($\sim$L7-T3). With the increasing coverage of UKIDSS we
can refine our selection techniques through additional
color-spectral type relationships in the near future.

Of the 36 objects with SDSS spectra, 19 have 2$\sigma$ detections of
non-zero proper motions from SDSS-2MASS, 10 of which have proper
motions above 0.2$^{\prime\prime}$yr$^{-1}$ (see, Table 1). There
are fewer 2$\sigma$ proper motion detections for objects without
spectra because they are, on average, further away. For SDSS-2MASS
match, a matching radius of 6$^{\prime\prime}$ might lead to the
loss of a small number hight proper motion objects (e.g. proper
motion larger than 1$^{\prime\prime}$yr$^{-1}$ and baseline longer
than 6 years). Some objects in Table 1,6 and 7 (online data) have
larger proper motions (also with large errors) but usually have a
shorter baseline (even less than a year) and these proper motions are not
very reliable. The errors in our proper motion measurements are
dominated by 2MASS positional uncertainties (especially for objects
with shorter baselines), however we have shown through a variety of
comparisons that our 2MASS-SDSS database proper motions are of
reasonable quality and can thus provide an additional tool to
identify large samples of L dwarfs in the SDSS sky. In the future, a
SDSS second epoch and surveys such as Pan-STARRS will offer an even
more powerful means to efficiently select late M and L dwarfs
through their proper motion.

It is clear that SDSS combined with 2MASS and now UKIDSS, offers a
powerful means to select large populations of L dwarfs using spectroscopy,
photometry and astrometry. As the sample of known L dwarfs grows we
can expect to reveal a broader range of inherent properties (e.g.
composition, mass, age, kinematics). Higher signal-to-noise and resolution
spectroscopic observations could be used to study such interesting
sub-populations e.g. by searching for the presence of lithium to directly
assess age and mass (Pavlenko et al. \cite{pav07}) and the use of
higher resolution cross correlation techniques to measure radial
velocities and space motions, yielding important kinematic information.

\begin{acknowledgements}
Funding for the SDSS and SDSS-II has been provided by the Alfred P.
Sloan Foundation, the Participating Institutions, the National
Science Foundation, the U.S. Department of Energy, the National
Aeronautics and Space Administration, the Japanese Monbukagakusho,
the Max Planck Society, and the Higher Education Funding Council for
England. The SDSS Web Site is http://www.sdss.org/.

The SDSS is managed by the Astrophysical Research Consortium for the
Participating Institutions. The Participating Institutions are the
American Museum of Natural History, Astrophysical Institute Potsdam,
University of Basel, University of Cambridge, Case Western Reserve
University, University of Chicago, Drexel University, Fermilab, the
Institute for Advanced Study, the Japan Participation Group, Johns
Hopkins University, the Joint Institute for Nuclear Astrophysics,
the Kavli Institute for Particle Astrophysics and Cosmology, the
Korean Scientist Group, the Chinese Academy of Sciences (LAMOST),
Los Alamos National Laboratory, the Max-Planck-Institute for
Astronomy (MPIA), the Max-Planck-Institute for Astrophysics (MPA),
New Mexico State University, Ohio State University, University of
Pittsburgh, University of Portsmouth, Princeton University, the
United States Naval Observatory, and the University of Washington.

This work is based in part on data obtained as part of the UKIRT
Infrared Deep Sky Survey. This publication makes use of data
products from the Two Micron All Sky Survey. This research has made
use of the VizieR catalogue access tool, CDS, Strasbourg, France.
Research has benefitted from the M, L, and T dwarf compendium housed
at DwarfArchives.org and maintained by Chris Gelino, Davy
Kirkpatrick, and Adam Burgasser. This work was part supported by the
Natural Science Foundation of China under Grant Nos 10521001,
10433030, 10503010 and the CAS Research Fellowship for International
Young Researchers.

\end{acknowledgements}


\clearpage

\Online

\begin{appendix} 

\longtab{7}{
\begin{landscape}
\begin{longtable}{c c c c c c c c l l}
\caption{\label{table7} SDSS and 2MASS photometry of 129 ultra-cool dwarf candidates}\\
\hline\hline
SDSS Name & SDSS \emph{r} & SDSS \emph{i} & SDSS \emph{z} & 2MASS \emph{J} & 2MASS \emph{H} & 2MASS \emph{K} & Proper Motion & Proper Motion & Sp. Type  \\
&  &  &  &  & & & ($^{\prime\prime}$yr$^{-1}$) & Angle & by Colors    \\
\hline
\endfirsthead
\caption{continued.}\\
\hline\hline
SDSS Name & SDSS \emph{r} & SDSS \emph{i} & SDSS \emph{z} & 2MASS \emph{J} & 2MASS \emph{H} & 2MASS \emph{K} & Proper Motion & Proper Motion & Sp. Type  \\
&  &  &  &  & & & ($^{\prime\prime}$yr$^{-1}$) & Angle & by Colors    \\
\hline
\endhead
\hline
\endfoot
SDSS J073813.07+155304.7 & 23.76$\pm$0.42 & 21.26$\pm$0.08 & 19.25$\pm$0.06 & 16.89$\pm$0.20 & 16.07$\pm$0.21 & 15.37$\pm$0.17 &0.04$\pm$0.04 &184 ...& L1 \\
SDSS J074151.17+275837.6 & 23.60$\pm$0.56 & 21.51$\pm$0.15 & 19.39$\pm$0.07 & 17.31$\pm$0.18 & 16.67$\pm$0.21 & 15.95$\pm$0.23 &0.05$\pm$0.08 &315 ...& L0 \\
SDSS J074838.61+174332.9 & 23.89$\pm$0.46 & 21.35$\pm$0.07 & 19.06$\pm$0.05 & 16.27$\pm$0.11 & 15.18$\pm$0.09 & 14.42$\pm$0.09 &0.07$\pm$0.02 &264$\pm$14& L7 \\
SDSS J075635.25+363033.6 & 22.26$\pm$0.20 & 19.85$\pm$0.06 & 17.07$\pm$0.01 & 15.24$\pm$0.04 & 14.61$\pm$0.05 & 14.09$\pm$0.05 &0.11$\pm$0.11 &149 ...& L2.5 \\
SDSS J075752.70+091410.0 & 22.86$\pm$0.27 & 20.62$\pm$0.05 & 18.57$\pm$0.03 & 15.86$\pm$0.08 & 14.83$\pm$0.07 & 14.09$\pm$0.06 &0.14$\pm$0.02 &249$\pm$~~8& L4 \\
SDSS J075923.05+462007.4 & 24.18$\pm$0.66 & 21.44$\pm$0.12 & 19.30$\pm$0.06 & 17.00$\pm$0.16 & 16.09$\pm$0.13 & 15.86$\pm$0.23 &0.23$\pm$0.25 &297 ...& L2 \\
SDSS J080020.39+360627.1 & 23.61$\pm$0.43 & 21.24$\pm$0.08 & 19.23$\pm$0.06 & 16.64$\pm$0.13 & 15.93$\pm$0.16 & 15.62$\pm$0.19 &0.37$\pm$0.25 &141$\pm$42& L2.5 \\
SDSS J080138.61+372205.8 & 23.00$\pm$0.23 & 20.59$\pm$0.08 & 18.21$\pm$0.02 & 16.83$\pm$0.22 & 15.84$\pm$0.22 & 15.45$\pm$0.19 &0.22$\pm$0.13 &254$\pm$35& ... \\
SDSS J080252.73+051058.3 & 23.59$\pm$0.64 & 20.72$\pm$0.09 & 18.67$\pm$0.06 & 16.16$\pm$0.11 & 15.30$\pm$0.09 & 14.91$\pm$0.14 &0.06$\pm$0.04 &270$\pm$44& L2 \\
SDSS J081215.88+504758.3 & 23.15$\pm$0.32 & 20.71$\pm$0.05 & 18.69$\pm$0.04 & 16.37$\pm$0.13 & 15.70$\pm$0.18 & 15.28$\pm$0.17 &0.24$\pm$0.05 &232$\pm$12& L0.5 \\
SDSS J081409.11+281909.6 & 24.25$\pm$0.74 & 21.69$\pm$0.18 & 19.58$\pm$0.10 & 17.45$\pm$0.23 & 16.47$\pm$0.21 & 15.79$\pm$0.18 &0.17$\pm$0.07 &246$\pm$23& L0.5 \\
SDSS J081825.78+374103.1 & 23.32$\pm$0.27 & 21.70$\pm$0.10 & 19.43$\pm$0.06 & 16.94$\pm$0.20 & 16.01$\pm$0.19 & 15.44$\pm$0.18 &0.19$\pm$0.10 &~~65$\pm$32& L3.5 \\
SDSS J081843.64+175645.7 & 23.14$\pm$0.34 & 21.61$\pm$0.15 & 19.44$\pm$0.09 & 17.01$\pm$0.19 & 16.02$\pm$0.19 & 15.45$\pm$0.17 &0.07$\pm$0.05 &349$\pm$47& L2.5 \\
SDSS J081905.47+493118.5 & 23.75$\pm$0.54 & 21.23$\pm$0.09 & 19.19$\pm$0.06 & 17.09$\pm$0.23 & 16.51$\pm$0.31 & 15.56$\pm$0.21 & ...  & ~~~~~... & L0 \\
SDSS J081951.43+543155.1 & 23.71$\pm$0.50 & 21.02$\pm$0.08 & 18.97$\pm$0.06 & 16.97$\pm$0.19 & 15.91$\pm$0.14 & 15.45$\pm$0.20 &0.13$\pm$0.06 &179$\pm$30& L1 \\
SDSS J082059.29+280648.0 & 23.22$\pm$0.36 & 21.22$\pm$0.09 & 19.12$\pm$0.06 & 16.92$\pm$0.21 & 15.89$\pm$0.19 & 15.40$\pm$0.16 &0.08$\pm$0.10 &240 ...& L1 \\
SDSS J082213.75+510120.2 & 23.46$\pm$0.43 & 20.85$\pm$0.07 & 18.81$\pm$0.05 & 16.57$\pm$0.12 & 15.66$\pm$0.12 & 15.17$\pm$0.12 & ...  & ~~~~~... & L0.5 \\
SDSS J083301.44+445107.6 & 23.45$\pm$0.43 & 21.19$\pm$0.08 & 19.12$\pm$0.05 & 16.54$\pm$0.14 & 15.96$\pm$0.19 & 15.42$\pm$0.20 &0.56$\pm$0.22 &152$\pm$23& L2 \\
SDSS J084537.80+293137.2 & 23.74$\pm$0.42 & 21.47$\pm$0.11 & 19.42$\pm$0.06 & 16.74$\pm$0.12 & 15.97$\pm$0.13 & 15.14$\pm$0.12 &0.07$\pm$0.04 &124$\pm$33& L2.5 \\
SDSS J084737.45+462443.8 & 23.16$\pm$0.50 & 20.92$\pm$0.09 & 18.87$\pm$0.07 & 16.88$\pm$0.17 & 16.03$\pm$0.19 & 15.78$\pm$0.23 &0.02$\pm$0.17 &~~76 ...& ...\\
SDSS J084750.21+510846.3 & 23.08$\pm$0.44 & 20.83$\pm$0.08 & 18.79$\pm$0.04 & 16.66$\pm$0.14 & 16.00$\pm$0.15 & 15.24$\pm$0.15 &0.20$\pm$0.20 &312$\pm$83& L0 \\
SDSS J084838.93+484025.4 & 23.07$\pm$0.27 & 21.04$\pm$0.07 & 19.03$\pm$0.05 & 16.59$\pm$0.13 & 15.77$\pm$0.13 & 15.29$\pm$0.15 &0.55$\pm$0.10 &106$\pm$11& L2 \\
SDSS J085048.96+173210.9 & 23.33$\pm$0.30 & 20.89$\pm$0.06 & 18.88$\pm$0.05 & 16.57$\pm$0.12 & 15.93$\pm$0.15 & 15.23$\pm$0.11 &0.15$\pm$0.03 &354$\pm$12& L0.5 \\
SDSS J085117.58+583226.6 & 24.25$\pm$0.76 & 21.59$\pm$0.13 & 19.53$\pm$0.09 & 17.07$\pm$0.19 & 16.45$\pm$0.23 & 15.84$\pm$0.23 &0.03$\pm$0.07 &257 ...& L1.5 \\
SDSS J085119.69+104348.0 & 24.36$\pm$0.65 & 21.79$\pm$0.11 & 19.58$\pm$0.05 & 16.79$\pm$0.15 & 15.78$\pm$0.15 & 15.09$\pm$0.12 &0.23$\pm$0.03 &309$\pm$~~7& L5 \\
SDSS J085711.43+415928.6 & 23.72$\pm$0.53 & 21.46$\pm$0.11 & 19.39$\pm$0.07 & 17.37$\pm$0.24 & 16.18$\pm$0.20 & 15.73$\pm$0.22 &0.26$\pm$0.10 &129$\pm$21& L2 \\
SDSS J090308.17+165935.5 & 24.87$\pm$0.59 & 21.61$\pm$0.10 & 19.38$\pm$0.07 & 16.49$\pm$0.11 & 15.78$\pm$0.16 & 15.49$\pm$0.19 &0.11$\pm$0.04 &129$\pm$20& L6 \\
SDSS J090435.90+322918.7 & 23.71$\pm$0.52 & 21.05$\pm$0.07 & 19.02$\pm$0.05 & 16.97$\pm$0.17 & 16.22$\pm$0.20 & 15.45$\pm$0.20 &0.38$\pm$0.06 &294$\pm$10& ...\\
SDSS J090546.54+562311.9 & 23.01$\pm$0.36 & 20.28$\pm$0.05 & 18.24$\pm$0.03 & 15.40$\pm$0.05 & 14.28$\pm$0.04 & 13.73$\pm$0.04 & ...  & ~~~~~... & L4 \\
SDSS J091428.64+230541.2 & 24.66$\pm$0.72 & 21.26$\pm$0.08 & 19.14$\pm$0.05 & 16.62$\pm$0.10 & 15.53$\pm$0.08 & 14.90$\pm$0.08 &0.05$\pm$0.02 &311$\pm$22& L3 \\
SDSS J091811.89+390216.7 & 23.77$\pm$0.62 & 22.07$\pm$0.21 & 19.65$\pm$0.10 & 17.02$\pm$0.21 & 16.15$\pm$0.21 & 15.57$\pm$0.22 &0.07$\pm$0.10 &234 ...& L5.5 \\
SDSS J091816.02+481300.4 & 23.02$\pm$0.23 & 20.75$\pm$0.05 & 18.74$\pm$0.05 & 16.43$\pm$0.12 & 15.52$\pm$0.12 & 15.30$\pm$0.14 &0.14$\pm$0.06 &~~71$\pm$27& L0.5 \\
SDSS J092752.44+572932.5 & 23.90$\pm$0.51 & 21.37$\pm$0.09 & 19.36$\pm$0.07 & 16.65$\pm$0.15 & 15.46$\pm$0.13 & 14.84$\pm$0.11 & ... & ~~~~~... & L3.5 \\
SDSS J092819.74+180510.8 & 23.64$\pm$0.35 & 22.20$\pm$0.15 & 19.98$\pm$0.08 & 17.23$\pm$0.21 & 16.65$\pm$0.26 & 15.90$\pm$0.21 &0.25$\pm$0.05 &289$\pm$11& L3.5 \\
SDSS J093100.69+605539.6 & 23.64$\pm$0.58 & 21.10$\pm$0.10 & 19.00$\pm$0.07 & 16.48$\pm$0.11 & 15.69$\pm$0.10 & 15.30$\pm$0.13 &0.44$\pm$0.15 &280$\pm$20& L2.5 \\
SDSS J093204.02+345937.0 & 25.18$\pm$0.73 & 21.97$\pm$0.22 & 19.92$\pm$0.10 & 17.01$\pm$0.18 & 16.22$\pm$0.19 & 15.46$\pm$0.21 &0.23$\pm$0.07 &149$\pm$16& L4.5 \\
SDSS J093956.05+242658.7 & 24.85$\pm$0.64 & 22.07$\pm$0.23 & 19.69$\pm$0.10 & 16.97$\pm$0.22 & 15.93$\pm$0.22 & 15.53$\pm$0.21 &0.36$\pm$0.05 &244$\pm$~~8& L6.5 \\
SDSS J094146.38+215843.5 & 25.07$\pm$0.58 & 21.78$\pm$0.11 & 19.67$\pm$0.09 & 16.94$\pm$0.18 & 15.92$\pm$0.18 & 15.55$\pm$0.17 &0.09$\pm$0.04 &244$\pm$26& L4 \\
SDSS J094427.33+641037.3 & 22.96$\pm$0.33 & 20.95$\pm$0.08 & 18.93$\pm$0.06 & 16.67$\pm$0.16 & 15.88$\pm$0.17 & 15.46$\pm$0.18 &0.09$\pm$0.06 &296$\pm$38& L0 \\
SDSS J094429.58+465254.7 & 22.84$\pm$0.32 & 21.26$\pm$0.11 & 19.16$\pm$0.06 & 17.26$\pm$0.21 & 16.58$\pm$0.23 & 16.09$\pm$0.25 &0.06$\pm$0.16 &322 ...& ...\\
SDSS J095154.19+282040.8 & 23.10$\pm$0.30 & 21.22$\pm$0.08 & 19.21$\pm$0.06 & 16.56$\pm$0.16 & 16.06$\pm$0.15 & 15.52$\pm$0.18 &0.07$\pm$0.05 &298$\pm$44& L2 \\
SDSS J095932.74+452330.5 & 23.73$\pm$0.51 & 21.04$\pm$0.08 & 19.01$\pm$0.05 & 15.88$\pm$0.07 & 14.76$\pm$0.07 & 13.67$\pm$0.04 &0.19$\pm$0.05 &221$\pm$15& L7.5 \\
SDSS J100132.26+492819.9 & 23.05$\pm$0.25 & 20.74$\pm$0.05 & 18.72$\pm$0.04 & 16.68$\pm$0.13 & 15.95$\pm$0.14 & 15.56$\pm$0.18 &0.23$\pm$0.09 &276$\pm$23& ...\\
SDSS J100317.30+331922.0 & 24.32$\pm$0.79 & 21.60$\pm$0.15 & 19.23$\pm$0.07 & 16.74$\pm$0.14 & 15.73$\pm$0.11 & 15.23$\pm$0.14 &0.05$\pm$0.04 &~~75$\pm$56& L4 \\
SDSS J100633.74+363919.5 & 23.89$\pm$0.39 & 21.60$\pm$0.08 & 19.49$\pm$0.05 & 17.10$\pm$0.22 & 15.98$\pm$0.16 & 15.32$\pm$0.19 &0.12$\pm$0.07 &184$\pm$32& L2.5 \\
SDSS J101134.72+501400.7 & 23.79$\pm$0.49 & 21.93$\pm$0.15 & 19.90$\pm$0.10 & 17.60$\pm$0.29 & 16.28$\pm$0.20 & 15.57$\pm$0.17 &0.70$\pm$0.22 &259$\pm$18& L2 \\
SDSS J101439.66+252511.3 & 24.45$\pm$0.59 & 21.14$\pm$0.07 & 19.10$\pm$0.05 & 17.24$\pm$0.23 & 16.19$\pm$0.21 & 15.83$\pm$0.24 &0.09$\pm$0.05 &215$\pm$35& L0 \\
SDSS J101951.13+044944.1 & 24.10$\pm$0.57 & 21.80$\pm$0.15 & 19.48$\pm$0.07 & 16.85$\pm$0.18 & 16.18$\pm$0.20 & 15.54$\pm$0.23 &0.57$\pm$0.31 &240$\pm$33& L4 \\
SDSS J102517.57+285113.6 & 23.64$\pm$0.54 & 20.95$\pm$0.07 & 18.92$\pm$0.05 & 16.54$\pm$0.10 & 15.83$\pm$0.12 & 15.31$\pm$0.12 &0.11$\pm$0.03 &288$\pm$16& L1.5 \\
SDSS J102546.97+151126.3 & 24.29$\pm$0.70 & 21.52$\pm$0.15 & 19.47$\pm$0.07 & 16.99$\pm$0.15 & 16.06$\pm$0.17 & 15.42$\pm$0.15 &0.17$\pm$0.04 &276$\pm$13& L2 \\
SDSS J102935.23+062029.6 & 24.51$\pm$0.55 & 22.47$\pm$0.19 & 19.37$\pm$0.05 & 16.87$\pm$0.22 & 16.09$\pm$0.21 & 15.02$\pm$0.16 &0.15$\pm$0.14 &212$\pm$75& L8 \\
SDSS J102939.69+571544.3 & 24.35$\pm$0.52 & 21.62$\pm$0.10 & 19.17$\pm$0.06 & 16.70$\pm$0.13 & 15.46$\pm$0.11 & 14.99$\pm$0.09 &0.82$\pm$0.07 &~~95$\pm$5& L4.5 \\
SDSS J103908.17+244044.1 & 23.58$\pm$0.34 & 21.73$\pm$0.11 & 19.41$\pm$0.06 & 16.76$\pm$0.14 & 15.72$\pm$0.13 & 15.04$\pm$0.11 &0.22$\pm$0.03 &236$\pm$~~7& L5 \\
SDSS J104808.29+544715.1 & 23.50$\pm$0.40 & 21.86$\pm$0.14 & 19.46$\pm$0.07 & 16.73$\pm$0.19 & 15.86$\pm$0.20 & 15.43$\pm$0.19 &0.38$\pm$0.11 &295$\pm$17& L6.5 \\
SDSS J104814.75+135833.3 & 23.78$\pm$0.44 & 21.94$\pm$0.15 & 19.43$\pm$0.07 & 16.90$\pm$0.13 & 16.01$\pm$0.14 & 15.34$\pm$0.14 &0.17$\pm$0.04 &197$\pm$14& L5.5 \\
SDSS J105204.75+172241.2 & 23.14$\pm$0.34 & 21.27$\pm$0.10 & 19.21$\pm$0.06 & 16.71$\pm$0.11 & 16.04$\pm$0.14 & 15.53$\pm$0.16 &0.31$\pm$0.03 &158$\pm$~~6& L1.5 \\
SDSS J105254.02+584951.2 & 23.96$\pm$0.37 & 21.44$\pm$0.08 & 19.26$\pm$0.04 & 16.44$\pm$0.12 & 15.44$\pm$0.13 & 14.87$\pm$0.10 &0.09$\pm$0.05 &~~54$\pm$37& L5 \\
SDSS J110827.31+083801.8 & 23.74$\pm$0.71 & 22.39$\pm$0.34 & 19.26$\pm$0.07 & 16.58$\pm$0.16 & 15.50$\pm$0.11 & 15.03$\pm$0.16 &0.34$\pm$0.08 &217$\pm$14& L8.5 \\
SDSS J111501.36+160701.5 & 22.94$\pm$0.37 & 20.71$\pm$0.07 & 18.59$\pm$0.05 & 16.40$\pm$0.12 & 15.22$\pm$0.09 & 14.56$\pm$0.11 &0.34$\pm$0.08 &248$\pm$14& L1 \\
SDSS J111802.89+060703.6 & 24.09$\pm$0.53 & 21.42$\pm$0.10 & 19.35$\pm$0.05 & 17.00$\pm$0.18 & 16.27$\pm$0.21 & 15.66$\pm$0.26 &0.28$\pm$0.11 &133$\pm$23& L1.5 \\
SDSS J111910.46+055248.4 & 23.39$\pm$0.33 & 21.67$\pm$0.11 & 19.60$\pm$0.06 & 16.76$\pm$0.16 & 15.48$\pm$0.11 & 15.03$\pm$0.15 &0.07$\pm$0.16 &109 ...& L5 \\
SDSS J112012.95+212520.4 & 23.50$\pm$0.42 & 21.65$\pm$0.11 & 19.53$\pm$0.06 & 16.90$\pm$0.17 & 15.79$\pm$0.14 & 15.34$\pm$0.14 &0.13$\pm$0.04 &198$\pm$16& L4 \\
SDSS J112722.94$-$003714.4 & 23.53$\pm$0.33 & 21.34$\pm$0.08 & 19.24$\pm$0.05 & 16.81$\pm$0.18 & 16.12$\pm$0.26 & 15.35$\pm$0.20 & ...  & ~~~~~... & L1.5 \\
SDSS J113022.46+122751.6 & 23.49$\pm$0.48 & 21.69$\pm$0.13 & 19.64$\pm$0.10 & 17.05$\pm$0.18 & 16.47$\pm$0.22 & 15.98$\pm$0.24 &0.16$\pm$0.11 &358$\pm$41& L2 \\
SDSS J113639.67+485240.3 & 23.54$\pm$0.31 & 20.91$\pm$0.05 & 18.86$\pm$0.03 & 16.16$\pm$0.10 & 15.27$\pm$0.11 & 14.58$\pm$0.08 &0.20$\pm$0.06 &307$\pm$17& L3.5 \\
SDSS J114103.28+632805.9 & 23.57$\pm$0.41 & 21.50$\pm$0.12 & 19.46$\pm$0.07 & 16.72$\pm$0.13 & 15.94$\pm$0.13 & 15.62$\pm$0.23 &0.45$\pm$0.18 &308$\pm$23& L3 \\
SDSS J114302.72+190541.9 & 24.77$\pm$0.62 & 22.76$\pm$0.26 & 19.77$\pm$0.08 & 16.77$\pm$0.18 & 15.80$\pm$0.15 & 15.01$\pm$0.14 &0.26$\pm$0.03 &172$\pm$~~8& L9 \\
SDSS J114807.23+390106.9 & 24.12$\pm$0.66 & 21.23$\pm$0.08 & 19.19$\pm$0.05 & 16.92$\pm$0.18 & 15.96$\pm$0.17 & 15.52$\pm$0.17 &0.20$\pm$0.06 &323$\pm$18& L1.5 \\
SDSS J115017.36+512502.4 & 24.28$\pm$0.60 & 22.02$\pm$0.18 & 19.98$\pm$0.11 & 16.87$\pm$0.20 & 16.03$\pm$0.18 & 15.08$\pm$0.12 &0.15$\pm$0.12 &221$\pm$51& L6 \\
SDSS J115058.98+440917.2 & 24.06$\pm$0.58 & 21.38$\pm$0.09 & 19.37$\pm$0.06 & 17.05$\pm$0.19 & 16.15$\pm$0.23 & 15.49$\pm$0.15 &0.14$\pm$0.06 &272$\pm$26& L1 \\
SDSS J115722.81+264119.6 & 23.71$\pm$0.47 & 21.06$\pm$0.07 & 18.89$\pm$0.04 & 16.71$\pm$0.12 & 15.91$\pm$0.15 & 15.02$\pm$0.11 &0.06$\pm$0.03 &263$\pm$25& L1 \\
SDSS J115820.75+043501.7 & 22.00$\pm$0.20 & 20.36$\pm$0.07 & 17.95$\pm$0.04 & 15.61$\pm$0.06 & 14.68$\pm$0.06 & 14.44$\pm$0.06 &0.86$\pm$0.86 &316 ...& L3 \\
SDSS J120136.14+135005.9 & 22.76$\pm$0.17 & 20.66$\pm$0.05 & 18.63$\pm$0.04 & 16.88$\pm$0.13 & 16.14$\pm$0.17 & 15.74$\pm$0.16 &0.03$\pm$0.04 &~~94 ...& ...\\
SDSS J120337.00+445333.4 & 23.85$\pm$0.45 & 21.74$\pm$0.12 & 19.52$\pm$0.06 & 17.43$\pm$0.25 & 16.55$\pm$0.25 & 15.56$\pm$0.18 &0.24$\pm$0.08 &281$\pm$20& L0.5 \\
SDSS J121846.56+410016.0 & 24.87$\pm$0.71 & 21.74$\pm$0.13 & 19.61$\pm$0.07 & 16.74$\pm$0.16 & 15.58$\pm$0.13 & 15.13$\pm$0.15 &0.05$\pm$0.05 &118 ...& L5.5 \\
SDSS J122218.47+364348.4 & 22.61$\pm$0.17 & 20.21$\pm$0.03 & 18.15$\pm$0.03 & 15.97$\pm$0.08 & 15.27$\pm$0.10 & 14.85$\pm$0.09 &0.27$\pm$0.04 &~~92$\pm$~~9& L0 \\
SDSS J122449.44+502154.1 & 23.98$\pm$0.45 & 21.62$\pm$0.10 & 19.14$\pm$0.06 & 16.53$\pm$0.13 & 15.66$\pm$0.11 & 14.86$\pm$0.10 &0.48$\pm$0.05 &204$\pm$~~6& L6 \\
SDSS J123256.67+484417.0 & 22.21$\pm$0.18 & 20.54$\pm$0.06 & 18.54$\pm$0.03 & 16.47$\pm$0.18 & 15.59$\pm$0.14 & 15.32$\pm$0.20 &0.44$\pm$0.11 &254$\pm$15& L0 \\
SDSS J124151.85+561541.6 & 23.44$\pm$0.42 & 20.86$\pm$0.07 & 18.81$\pm$0.04 & 16.42$\pm$0.11 & 15.61$\pm$0.12 & 14.95$\pm$0.11 &0.04$\pm$0.04 &314$\pm$77& L2 \\
SDSS J124609.07+294200.1 & 24.07$\pm$0.48 & 21.30$\pm$0.08 & 19.26$\pm$0.05 & 16.97$\pm$0.19 & 16.12$\pm$0.21 & 15.65$\pm$0.24 &0.02$\pm$0.06 &331 ...& L1 \\
SDSS J124655.54+535342.8 & 23.89$\pm$0.47 & 21.31$\pm$0.08 & 19.24$\pm$0.05 & 16.37$\pm$0.11 & 15.57$\pm$0.12 & 14.98$\pm$0.12 &0.17$\pm$0.09 &160$\pm$33& L4 \\
SDSS J125002.99+484834.3 & 23.99$\pm$0.47 & 21.93$\pm$0.16 & 19.88$\pm$0.09 & 17.30$\pm$0.26 & 15.82$\pm$0.14 & 15.25$\pm$0.14 &0.37$\pm$0.15 &263$\pm$24& L4 \\
SDSS J125137.76+462026.0 & 22.94$\pm$0.21 & 20.59$\pm$0.05 & 18.50$\pm$0.03 & 15.83$\pm$0.08 & 14.75$\pm$0.07 & 14.48$\pm$0.10 &0.31$\pm$0.02 &172$\pm$~~4& L4 \\
SDSS J125410.69+382546.0 & 23.80$\pm$0.46 & 21.48$\pm$0.11 & 19.42$\pm$0.06 & 16.88$\pm$0.15 & 16.11$\pm$0.18 & 15.65$\pm$0.16 &0.06$\pm$0.04 &239$\pm$47& L2.5 \\
SDSS J125438.50+434657.2 & 23.70$\pm$0.48 & 22.22$\pm$0.22 & 19.47$\pm$0.07 & 16.81$\pm$0.15 & 15.35$\pm$0.10 & 14.81$\pm$0.09 &0.20$\pm$0.04 &257$\pm$11& L8.5 \\
SDSS J130449.88+010627.0 & 23.10$\pm$0.31 & 21.37$\pm$0.10 & 19.30$\pm$0.06 & 16.69$\pm$0.17 & 15.80$\pm$0.17 & 15.12$\pm$0.17 &0.83$\pm$0.24 &125$\pm$17& L3 \\
SDSS J131142.11+362923.9 & 23.08$\pm$0.23 & 20.49$\pm$0.04 & 18.31$\pm$0.03 & 15.55$\pm$0.05 & 14.75$\pm$0.06 & 14.14$\pm$0.05 &0.40$\pm$0.02 &279$\pm$~~3& L4 \\
SDSS J131218.20+284643.0 & 24.13$\pm$0.57 & 21.36$\pm$0.09 & 19.31$\pm$0.06 & 17.27$\pm$0.26 & 16.33$\pm$0.32 & 15.72$\pm$0.21 &0.10$\pm$0.09 &219$\pm$58& L0 \\
SDSS J132535.68+504007.0 & 24.45$\pm$0.54 & 21.56$\pm$0.08 & 19.38$\pm$0.05 & 16.93$\pm$0.23 & 15.86$\pm$0.23 & 15.46$\pm$0.23 &0.32$\pm$0.15 &288$\pm$27& L3 \\
SDSS J133316.06+374421.7 & 23.31$\pm$0.29 & 21.01$\pm$0.06 & 18.65$\pm$0.03 & 15.89$\pm$0.06 & 14.90$\pm$0.05 & 14.30$\pm$0.05 &0.12$\pm$0.04 &~~76$\pm$18& L6.5 \\
SDSS J134210.11+414023.8 & 23.65$\pm$0.35 & 21.63$\pm$0.11 & 19.56$\pm$0.07 & 16.86$\pm$0.13 & 15.87$\pm$0.12 & 15.47$\pm$0.14 &0.20$\pm$0.04 &229$\pm$12& L4 \\
SDSS J135640.46+140205.3 & 23.83$\pm$0.35 & 20.96$\pm$0.08 & 18.36$\pm$0.03 & 16.62$\pm$0.15 & 16.10$\pm$0.22 & 15.28$\pm$0.15 &0.03$\pm$0.06 &317 ...& L1 \\
SDSS J140058.03+234923.9 & 23.67$\pm$0.39 & 21.51$\pm$0.09 & 19.48$\pm$0.06 & 16.73$\pm$0.17 & 16.12$\pm$0.23 & 15.38$\pm$0.22 &0.06$\pm$0.06 &261 ...& L3 \\
SDSS J140318.98+243718.0 & 23.56$\pm$0.33 & 21.55$\pm$0.09 & 19.52$\pm$0.07 & 17.18$\pm$0.23 & 16.06$\pm$0.22 & 15.50$\pm$0.22 &0.03$\pm$0.06 &194 ...& L1.5 \\
SDSS J141118.49+294850.4 & 23.52$\pm$0.37 & 21.07$\pm$0.07 & 18.76$\pm$0.04 & 16.20$\pm$0.09 & 15.43$\pm$0.11 & 15.09$\pm$0.11 &0.28$\pm$0.05 &164$\pm$10& L3.5 \\
SDSS J141405.84+010710.4 & 23.40$\pm$0.38 & 21.77$\pm$0.15 & 19.63$\pm$0.09 & 16.74$\pm$0.20 & 15.74$\pm$0.19 & 15.25$\pm$0.20 & ...  & ~~~~~... & L5.5 \\
SDSS J142110.77+472834.3 & 23.92$\pm$0.39 & 20.40$\pm$0.03 & 18.33$\pm$0.02 & 16.16$\pm$0.09 & 15.33$\pm$0.10 & 14.98$\pm$0.13 &0.07$\pm$0.03 &240$\pm$22& L0 \\
SDSS J142404.53+184641.4 & 24.16$\pm$0.50 & 21.41$\pm$0.09 & 19.34$\pm$0.06 & 17.08$\pm$0.19 & 16.44$\pm$0.26 & 15.50$\pm$0.16 &0.01$\pm$0.07 &175 ...& L0.5 \\
SDSS J142527.14+300400.4 & 23.11$\pm$0.25 & 20.78$\pm$0.05 & 18.72$\pm$0.03 & 16.39$\pm$0.12 & 15.54$\pm$0.13 & 15.03$\pm$0.12 &0.02$\pm$0.04 &292 ...& L1.5 \\
SDSS J142612.86+313039.4 & 23.70$\pm$0.42 & 21.39$\pm$0.09 & 19.29$\pm$0.05 & 16.62$\pm$0.16 & 15.59$\pm$0.13 & 14.72$\pm$0.09 &0.19$\pm$0.04 &224$\pm$11& L4 \\
SDSS J143412.02+271729.9 & 23.98$\pm$0.62 & 21.41$\pm$0.11 & 19.29$\pm$0.07 & 17.38$\pm$0.24 & 16.33$\pm$0.21 & 15.70$\pm$0.21 &0.08$\pm$0.07 &~~96$\pm$68& L0 \\
SDSS J143636.98+465302.7 & 23.75$\pm$0.38 & 21.92$\pm$0.11 & 19.92$\pm$0.08 & 16.99$\pm$0.16 & 16.13$\pm$0.15 & 15.81$\pm$0.20 &0.28$\pm$0.07 &268$\pm$15& L4 \\
SDSS J145052.71+462024.8 & 23.87$\pm$0.45 & 21.68$\pm$0.10 & 19.67$\pm$0.09 & 16.77$\pm$0.15 & 15.93$\pm$0.17 & 15.26$\pm$0.14 &0.14$\pm$0.07 &305$\pm$32& L4 \\
SDSS J150651.27+553350.7 & 22.93$\pm$0.32 & 21.10$\pm$0.10 & 19.10$\pm$0.08 & 16.69$\pm$0.12 & 15.80$\pm$0.13 & 15.30$\pm$0.15 &0.08$\pm$0.11 &206 ...& L2 \\
SDSS J151110.91+434036.3 & 23.48$\pm$0.33 & 21.59$\pm$0.09 & 19.29$\pm$0.05 & 16.60$\pm$0.15 & 15.47$\pm$0.13 & 14.70$\pm$0.13 &0.24$\pm$0.04 &152$\pm$~~9& L5 \\
SDSS J152427.98+024210.1 & 22.76$\pm$0.29 & 21.21$\pm$0.12 & 19.16$\pm$0.06 & 16.98$\pm$0.20 & 16.39$\pm$0.24 & 15.35$\pm$0.17 & ... &  ~~~~~... & L0 \\
SDSS J152802.91+013949.6 & 23.06$\pm$0.33 & 21.38$\pm$0.11 & 19.27$\pm$0.12 & 16.65$\pm$0.14 & 15.75$\pm$0.15 & 15.26$\pm$0.17 & ... &  ~~~~~... & L3 \\
SDSS J153607.12+203032.6 & 24.01$\pm$0.53 & 21.81$\pm$0.14 & 19.73$\pm$0.10 & 17.26$\pm$0.21 & 16.17$\pm$0.20 & 15.77$\pm$0.24 &0.03$\pm$0.07 &243 ...& L2.5 \\
SDSS J153848.19+360337.5 & 24.05$\pm$0.44 & 21.68$\pm$0.11 & 19.56$\pm$0.05 & 16.64$\pm$0.15 & 15.76$\pm$0.13 & 15.29$\pm$0.14 &0.06$\pm$0.07 &286 ...& L5.5 \\
SDSS J153941.94+531131.0 & 23.53$\pm$0.44 & 21.44$\pm$0.12 & 19.40$\pm$0.07 & 16.86$\pm$0.16 & 15.95$\pm$0.17 & 15.19$\pm$0.18 &0.17$\pm$0.09 &204$\pm$34& L2 \\
SDSS J154038.76$-$001257.1 & 23.11$\pm$0.28 & 21.60$\pm$0.10 & 19.35$\pm$0.06 & 16.80$\pm$0.15 & 15.67$\pm$0.12 & 15.05$\pm$0.14 & ... & ~~~~~... & L4 \\
SDSS J154455.20+330145.1 & 22.96$\pm$0.23 & 20.54$\pm$0.05 & 18.33$\pm$0.02 & 15.55$\pm$0.06 & 14.52$\pm$0.06 & 13.94$\pm$0.05 &0.11$\pm$0.03 &~~25$\pm$15& L5 \\
SDSS J154623.27+333803.2 & 24.03$\pm$0.50 & 21.13$\pm$0.08 & 19.05$\pm$0.05 & 16.49$\pm$0.11 & 15.77$\pm$0.13 & 15.33$\pm$0.14 &0.08$\pm$0.07 &228$\pm$54& L2.5 \\
SDSS J155151.04+174216.7 & 23.34$\pm$0.24 & 21.10$\pm$0.06 & 19.08$\pm$0.06 & 16.70$\pm$0.14 & 15.83$\pm$0.15 & 15.13$\pm$0.12 &0.12$\pm$0.05 &320$\pm$24& L1.5 \\
SDSS J155702.82+121258.1 & 23.84$\pm$0.55 & 21.25$\pm$0.10 & 19.19$\pm$0.08 & 17.12$\pm$0.25 & 16.02$\pm$0.19 & 15.51$\pm$0.26 &0.62$\pm$0.05 &243$\pm$~~4& L0.5 \\
SDSS J160022.86+484132.8 & 23.53$\pm$0.44 & 21.20$\pm$0.10 & 19.06$\pm$0.05 & 16.27$\pm$0.09 & 15.43$\pm$0.12 & 15.01$\pm$0.14 &0.46$\pm$0.09 &102$\pm$11& L4.5 \\
SDSS J160835.64+120226.8 & 23.22$\pm$0.26 & 20.56$\pm$0.04 & 18.56$\pm$0.03 & 16.37$\pm$0.12 & 15.35$\pm$0.11 & 15.00$\pm$0.14 &0.07$\pm$0.03 &301$\pm$28& L0.5 \\
SDSS J160911.45+211658.7 & 24.02$\pm$0.64 & 21.53$\pm$0.14 & 19.50$\pm$0.08 & 16.96$\pm$0.21 & 15.97$\pm$0.20 & 14.87$\pm$0.11 &0.09$\pm$0.03 &~~30$\pm$22& L2 \\
SDSS J161231.01+483357.5 & 24.34$\pm$0.68 & 21.27$\pm$0.08 & 19.17$\pm$0.06 & 16.20$\pm$0.10 & 15.62$\pm$0.12 & 14.83$\pm$0.12 &0.11$\pm$0.06 &320$\pm$38& L5.5 \\
SDSS J161459.98+400435.1 & 22.60$\pm$0.19 & 20.95$\pm$0.08 & 18.93$\pm$0.04 & 16.57$\pm$0.12 & 15.84$\pm$0.15 & 15.01$\pm$0.12 &0.30$\pm$0.04 &296$\pm$~~8& L1 \\
SDSS J161655.06+190842.8 & 23.57$\pm$0.42 & 21.46$\pm$0.09 & 19.45$\pm$0.07 & 17.18$\pm$0.19 & 16.37$\pm$0.22 & 15.90$\pm$0.24 &0.06$\pm$0.05 &305$\pm$59& L0 \\
SDSS J163021.84$-$001801.6 & 24.80$\pm$0.69 & 22.58$\pm$0.24 & 19.42$\pm$0.06 & 16.25$\pm$0.12 & 15.55$\pm$0.15 & 15.24$\pm$0.19 & ... & ~~~~~... & L9.5 \\
SDSS J164939.25+425043.7 & 24.06$\pm$0.46 & 21.71$\pm$0.11 & 19.48$\pm$0.07 & 16.79$\pm$0.14 & 15.46$\pm$0.10 & 14.79$\pm$0.12 &0.18$\pm$0.05 &320$\pm$15& L5 \\
SDSS J165914.44+172642.3 & 24.22$\pm$0.48 & 21.73$\pm$0.10 & 19.69$\pm$0.06 & 16.69$\pm$0.16 & 15.54$\pm$0.13 & 15.15$\pm$0.14 &0.11$\pm$0.06 &107$\pm$32& L6 \\
SDSS J170418.24+744315.0 & 23.26$\pm$0.30 & 21.32$\pm$0.07 & 19.07$\pm$0.05 & 16.09$\pm$0.09 & 15.14$\pm$0.09 & 14.27$\pm$0.09 &0.13$\pm$0.04 &217$\pm$18& L7.5 \\
SDSS J171201.36+324456.6 & 23.99$\pm$0.44 & 21.29$\pm$0.08 & 19.25$\pm$0.07 & 17.14$\pm$0.17 & 16.26$\pm$0.18 & 15.96$\pm$0.24 &0.52$\pm$0.10 &272$\pm$11& L0 \\
SDSS J172545.59+640501.2 & 23.90$\pm$0.81 & 21.90$\pm$0.23 & 19.46$\pm$0.09 & 16.81$\pm$0.17 & 15.89$\pm$0.17 & 15.35$\pm$0.20 &0.61$\pm$0.33 &~~12$\pm$33& L6.5 \\
SDSS J172746.49+572247.6 & 22.53$\pm$0.21 & 20.21$\pm$0.04 & 18.21$\pm$0.03 & 15.83$\pm$0.08 & 14.96$\pm$0.09 & 14.69$\pm$0.11 &0.14$\pm$0.10 &~~33$\pm$51& L1.5 \\
SDSS J210515.30$-$003701.5 & 24.48$\pm$0.60 & 21.33$\pm$0.08 & 19.10$\pm$0.05 & 16.97$\pm$0.17 & 15.90$\pm$0.17 & 14.91$\pm$0.13 &0.07$\pm$0.07 &~~81$\pm$64& L1.5 \\
\end{longtable}
\end{landscape}
}

\end{appendix}

\end{document}